\documentclass[journal]{IEEEtran}

\usepackage{amsmath, amsfonts, amssymb, amsthm, amsbsy, amscd, bm} 
\usepackage{mathrsfs}
\usepackage{cite}
\usepackage{array}
\usepackage{rotating}
\usepackage{hyperref}
\usepackage{subfigure}

\usepackage{comment,array}
\usepackage{graphicx}
\usepackage[table]{xcolor}

\usepackage{mwe}
\usepackage[percent]{overpic}

\makeatletter
\def\hlinewd#1{%
  \noalign{\ifnum0=`}\fi\hrule \@height #1 \futurelet
   \reserved@a\@xhline}
\makeatother

\usepackage{amsmath,graphicx}
\usepackage{amsfonts, amssymb, amsbsy, amscd, bm} 
\usepackage{mathrsfs}
\usepackage{array}
\usepackage{booktabs}
\usepackage{tikz}
\usepackage{subfigure}
\usepackage{multicol}
\usepackage{multirow}
\usepackage{rotating}
\usepackage{arydshln}
\usepackage{pstricks}
\usepackage{enumerate}
\usepackage{color}
\usepackage{hyperref}
\usepackage{algorithmic}
\usepackage[lined,ruled,linesnumbered]{algorithm2e}
\usepackage{microtype}
\usepackage{booktabs}
\usepackage{float}

\makeatletter
\def\hlinewd#1{%
  \noalign{\ifnum0=`}\fi\hrule \@height #1 \futurelet
   \reserved@a\@xhline}
\makeatother

\newcommand*{\twoelementtable}[3][l]%
{%
    \renewcommand{\arraystretch}{0.8}%
    \begin{tabular}[t]{@{}#1@{}}%
        #2\tabularnewline
        #3%
    \end{tabular}%
}

\definecolor{mygreen1}{rgb}{0, .4, 0}
\definecolor{mygreen2}{rgb}{0, .8, 0}

\graphicspath{{figs/}}

\ifCLASSINFOpdf
\else 
\fi

\begin{document}
%

\title{Focus Quality Assessment of High-Throughput Whole Slide Imaging in Digital Pathology}

%
%
%

\author{Mahdi~S.~Hosseini,~\IEEEmembership{Member,~IEEE,},
        Yueyang~Zhang,~\IEEEmembership{Student member,~IEEE},
				Lyndon~Chan,~\IEEEmembership{Student member,~IEEE},				
        Konstantinos~N.~Plataniotis,~\IEEEmembership{Fellow,~IEEE},
				~Jasper~A.~Z.~Brawley-Hayes,
				and~Savvas~Damaskinos
\thanks{Copyright (c) 2018 IEEE. Personal use of this material is permitted. However, permission to use this material for any other purposes must be obtained from the IEEE by sending a request to pubs-permissions@ieee.org. 

M. S. Hosseini, Y. Zhang, L. Chan and K. N. Plataniotis are with The Edward S. Rogers Sr. Department of Electrical and Computer Engineering, University of Toronto, Toronto, ON M5S 3G4, Canada e-mail: \url{mahdi.hosseini@mail.utoronto.ca}.

M.~S.~Hosseini,~J.~A.~Z.~Brawley-Hayes,~are~S.~Damaskinos are with Huron Digital Pathology Inc., Waterloo, ON N2L 5V4, Canada.}}%

%
%

\markboth{This work has been submitted to the IEEE for possible publication}%
{Shell \MakeLowercase{\textit{et al.}}: Bare Demo of IEEEtran.cls for IEEE Journals}
%



\maketitle

\begin{abstract} 
One of the challenges facing the adoption of digital pathology workflows for clinical use is the need for automated quality control. As the scanners sometimes determine focus inaccurately, the resultant image blur deteriorates the scanned slide to the point of being unusable. Also, the scanned slide images tend to be extremely large when scanned at greater or equal 20X image resolution. Hence, for digital pathology to be clinically useful, it is necessary to use computational tools to quickly and accurately quantify the image focus quality and determine whether an image needs to be re-scanned. We propose a no-reference focus quality assessment metric specifically for digital pathology images, that operates by using a sum of even-derivative filter bases to synthesize a human visual system-like kernel, which is modeled as the inverse of the lens' point spread function. This kernel is then applied to a digital pathology image to modify high-frequency image information deteriorated by the scanner's optics and quantify the focus quality at the patch level. We show in several experiments that our method correlates better with ground-truth $z$-level data than other methods, and is more computationally efficient. We also extend our method to generate a local slide-level focus quality heatmap, which can be used for automated slide quality control, and demonstrate the utility of our method for clinical scan quality control by comparison with subjective slide quality scores.
\end{abstract}
%

%

\section{Introduction}\label{sec:intro}
\IEEEPARstart{C}{linical} pathology is witnessing a paradigm shift by transferring from glass tissue slides (observed by optical microscopy) to digital slides scanned with whole slide imaging (WSI) systems. WSI scanners generate high resolution (submicron) images that can aid in the diagnosis of disease by permitting convenient visualization and navigation of tissue slide images. In clinical pathology departments, hundreds to thousands of tissue slides are processed each day and the number of cases has been steadily increasing, creating a challenging workload for digital pathology \cite{sanchez2018graph}. Routine diagnosis of pathology slides requires high quality high-throughput images, which are directly affected by the dynamic environment of the physical optics and sensor electronics of the scanner \cite{pantanowitz2011review, farahani2015whole, bueno2016new}.

High quality images from the slides must be displayed in a consistent and reliable manner to assist pathologists in interpreting these images accurately and efficiently. Existing digital pathology solutions, however still face challenges. For example, many WSI scanners need a manual inspection of digital slides for focus quality control (FQC), which is a tedious and laborious task. In high-throughput scanning systems, which contain hundreds of slides for processing, making it impractical to perform a manual FQC check for each individual slide. A robust, highly reliable automated solution (“one-click” mode) eliminates the need for manual FQC checking, and makes the integrated imaging software user friendly. The computational complexity of any automated FQC assessment technique should also be compatible with widely available scanning platforms, without the requirement of custom hardware. It can be a cost effective solution for digital pathology departments.

In discrete computational modeling, there exist several no-reference (NR) focus quality assessment (FQA) (together called NR-FQA) metrics that can be applied in synthetic or natural imaging applications for sharpness assessment such as using gradient map \cite{bahrami2014fast, li2016image, li2016no, li2017no, liu2017quality}, contrast map \cite{liu2012image, guan2015no, gvozden2018blind}, phase coherency \cite{hassen2013image, leclaire2015no}, and deep learning solutions \cite{kang2014convolutional, yu2016cnn, yu2017shallow}. Please refer to \cite{hosseini2018focus} and the references therein for more information on the methodologies. The performance accuracy of these metrics is usually evaluated by comparison with the subjective scores of a test database where the blur level of the image is scored by an individual subject. This score is correlated with the objective scores provided by NR-FQA metric for accuracy measurement. Despite their effectiveness for synthetic blur assessment, the correlation accuracy of such methods is usually low for naturally blurred images, partly because such images are usually captured in different blurring types, various scales, and illumination conditions. Beyond such irregularities, high accuracy NR-FQAs in the literature are highly complex in terms of computational speed, which makes them impractical for real-time applications.

The general application of NR-FQA metrics developed for blurred images is different from out-of-focus blur in digital pathology. In fact, when it comes to the digital pathology, imaging is done in a controlled environment where (a) the magnification is consistent through the image, with most manufacturers using a similar pixel pitch, (b) image uniformity is controlled by the quality of illumination optics to uniformly distribute light at the tissue located at the focal plane of the objective lens, and (c) poor focusing is the dominant factor of image blur in scanned images while image blur caused by motion of slide in the direction of scan is comparatively very small. Various methods take advantage of the scan control environment to employ simplified approaches for NR-FQA development in digital pathology. To meet the needs for efficiency and fast computational speed, these methods employ statistical measurements for feature extraction that are related to absolute image blurriness. For detailed review of these methods please refer to \cite{redondo2012autofocus, lopez2013automated, vasconcelos2014no, jimenez2016image, koho2016image, cabazos2018automatic, hurtado2018focus, campanella2018towards} and references therein.


Analysis of blur using such methods is highly effective for calculating the relative difference between images where structural similarity is preserved across different focus levels, known as the in-depth Z-stacks \cite{bray2012workflow}. However, using statistical measurements for focus quality assessment of image patches at the whole slide level is a challenge. The diversity of both structures (tissue morphology patterns) and focus quality levels across different positions in WSIs makes NR-FQA metric development a hard problem to solve. In fact, very few methods exist in the literature to address NR-FQA so as to create a local quality map (like a heatmap) to reliably assess the in-focus vs out-of-focus regions in a WSI. Recent developments tend to solve this problem either by statistical regression of blur features \cite{lahrmann2013semantic, campanella2018towards, bray2018quality} or by developing a deep-learning solution with convolutional neural networks (CNNs) \cite{yang2018assessing}. Unfortunately, the lack of (a) generalization to a broad spectrum of tissue types and (b) computational efficiency are the main drawbacks of such methodologies. This makes the NR-FQA still an unresolved problem in digital pathology for focus quality assessment in WSI.

We propose an NR-FQA metric called FQPath \footnote{\url{https://github.com/mahdihosseini/FQPath}} which extracts focus-related features from a digital pathology image that requires no prior knowledge from the user for reliable analysis of the blur levels to (a) grade the image quality across different patches in a WSI and (b) grade the WSI scan quality across slides of different tissue types. The key idea behind our solution is the synthesis of a convolution filter using the MaxPol derivative kernel library \cite{HosseiniPltaniotis_MaxPol_TIP_2017, HosseiniPltaniotis_MaxPol_SIAM_2017}, that mimics the human visual system response for microscopy applications in order to equalize the frequency spectrum of the image for blur feature extraction. We use this feature to develop our NR-FQA metric which is a variant of the method detailed in a previous paper for general-application NR-FQA \cite{mahdi2018image}. The proposed NR-FQA metric i.e. FQPath extracts features with this synthesized kernel instead of fixed derivative kernels, and accuracy is measured by correlation with scanning z-level instead of subjective score.

In this paper, we make the following contributions to improving WSI in digital pathology:
\begin{itemize}
	\item We develop a novel method for measuring how close the specimen was to the object plane at the time of acquisition using a NR-FQA. This method is based on the human visual system’s characteristic equalization of spatial frequency expression in an image, which allows us to distinguish between subtle changes in sharpness as a result of defocusing.
	\item We leverage the NR-FQA metric to generate a focus quality heatmap of whole slide images by applying the metric on small image patches. We propose an inverse Gaussian mapping of the metric scores to further improve the usability of the heatmap.
	\item We introduce a database of 200 WSI scans of histology slides stained with H$\&$E containing a variety of tissue types. From these images, we create a subjective scoring protocol to grade the focus quality at the WSI level.
	\item We conduct two sets of experiments to (a) compare ten state-of-the-art NR-FQA metrics based on their correlation to actual z-level; and (b) provide a quality control tool to aid in rapid grading of the focus level of WSI scans and provide the means to define a pass/fail threshold based on subjective scoring of images.
\end{itemize}

The work accomplished in this paper is described in sections below as follows. In Section \ref{sec:hvs_kernel_synthesis} we introduce the concept of HVS convolution kernel synthesis based on out-of-focus imaging in optical microscopy. In Section \ref{sec:focus} we adopt the HVS convolution kernel and propose an NR-FQA metric that we apply on digital image patches. In Section \ref{sec:heatmap} we construct a focus quality heatmap of WSI scans while in Sections \ref{sec:experiment_1} and \ref{sec:experiment_heatmap} we describe two sets of experiments to validate the proposed approach of focus quality assessment of digital pathology images. Finally, conclusions and recommendations are presented in Section \ref{conclusion}.
\section{Human Visual Response-Microscopy Like Kernel Synthesis} \label{sec:hvs_kernel_synthesis}
In this section, we explain our approach to model the out-of-focus characteristics of a digital pathology scanner that is caused by focal depth offset between the tissue slide and the scanner's objective lens. In particular, we synthesize a human visual system (HVS)-like kernel for optical microscopy (called HVS-Microscopy, or HVS-M for short) from derivative basis filters. This HVS-M kernel is then used to partly correct the high-frequency fall-off in the scanned image's amplitude response caused by the microscope's optics being out-of-focus relative to the plane of the sample. The corrected image is used to extract focus quality-related features.

\subsection{Out-of-Focus in WSI Scanning System} \label{subsec_out_of_focus_WSI}
One common practice in whole slide imaging (WSI) to determine the focus map of a tissue specimen on a glass slide is to pre-measure the focus positions at several locations on the slide and interpolate a surface that passes through the measured focus points. This is accomplished with a fast low-resolution ($\leq$ 1X) preview scan of the whole tissue slide.  A region or regions-of-interest (ROIs) are defined to be scanned at high-resolution ($\geq$ 20X). Within these regions (or a single region), a set of focus points are defined on the tissue. At each point, the scanner will perform a through-focus scan, typically a volume image, from which one lateral dimension may be discarded. The resulting xz-scans or xy-scans (one from each focus point) are run through an algorithm to determine the row or column of the image corresponding to the sample being at best focus. Since the image capture was synchronized with a stage encoder, the focus positions can be determined from these images. Using the known ($x_i$,$y_i$) coordinates of each measurement of $z_i$, a surface can be interpolated through all measured points. This interpolated surface is used to compute the optimal $z$ position for each frame during final high-resolution image acquisition.

During acquisition, errors may occur, which can lead to some or all parts of the image being out of focus. Some of these factors may be:
\begin{itemize}
\item	thermal effects in the mechanical system, which affect the repeatability of the position of the sample in relation to the indicated encoder value
\item	internally or externally generated vibration
\item	errors in the focus determination algorithm
\item	errors in the generation of the focus map
\item	insufficient number of focus points to define an accurate focus map
\item	errors in tissue preparation, such as folds or lifting off the slide surface
\end{itemize}
These factors cause the sample to pass either above or below the object plane of the objective lens, which leads to a blurred image at the camera sensor. In our approach, we attempt to quantify the image quality deterioration based on defocus, which may be caused by any of these factors.

In optical microscopy, the point spread function (PSF) describes a general diffraction model of the optical spreading, for both in-focus and out-of-focus behavior, that results when a focused imaging lens is (by definition) aligned with the target specimen in the lateral plane, i.e. (X,Y) coordinates, but not necessarily along the optical axis, i.e. Z coordinate. For different focal depth offsets (Z) from the optimal value, a different lateral characteristic will be observed. We chose to model the digital pathology scanner optical characteristic using the Born \& Wolf model \cite{born2013principles}.

\begin{align}
h_{PSF}(r,z) = \left|C\int_0^1{J_0(k\frac{\texttt{NA}}{n}r\rho)e^{-\frac{1}{2}ik\rho^2z(\frac{\texttt{NA}}{n})^2}\rho d\rho}\right|^2
\label{equation_PSF_1}
\end{align}
where,\\
$r$ - the radial distance from the optical axis along the lateral plane ($r=\sqrt{x^2+y^2}$, where $x$ and $y$ are the lateral Cartesian coordinates)\\
$z$ - the distance between the imaging plane and the in-focus position along the optical axis\\
$C$ - a normalization constant\\
$J_{0}$ - zero-order Bessel function of the first kind\\
$k$ - angular wavenumber of the lightsource ($k=\frac{2\pi}{\lambda}$)\\
$\texttt{NA}$ - the numerical aperture of the objective lens\\
$n$ - the refractive index of the medium\\
$i$ - the imaginary number\\
$\rho$ - the normalized coordinate in the exit pupil.

\begin{figure}[htp]
\centerline{
\subfigure[$x=0$]{\includegraphics[height=0.12\textwidth]{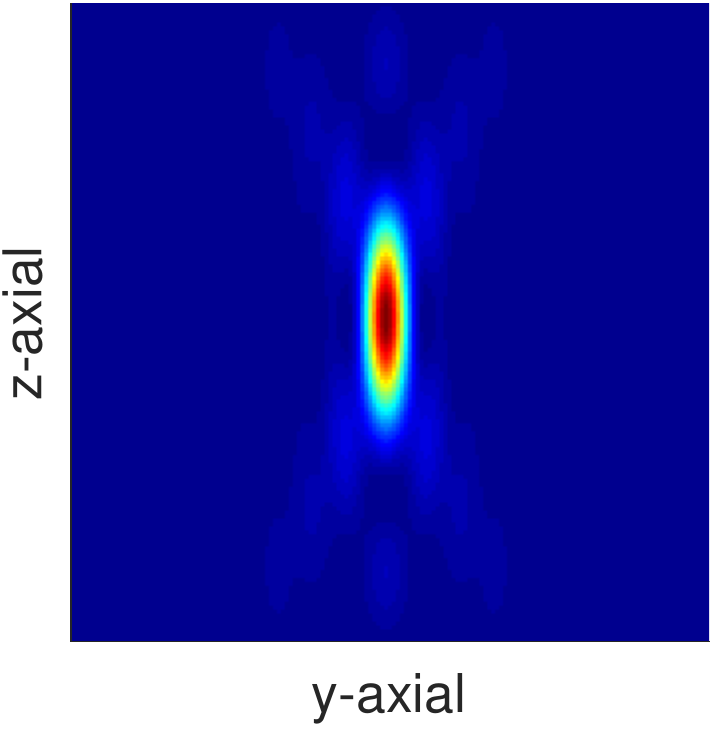}\label{PSF_responses_xCut_0}}
\subfigure[$x=4$]{\includegraphics[height=0.12\textwidth]{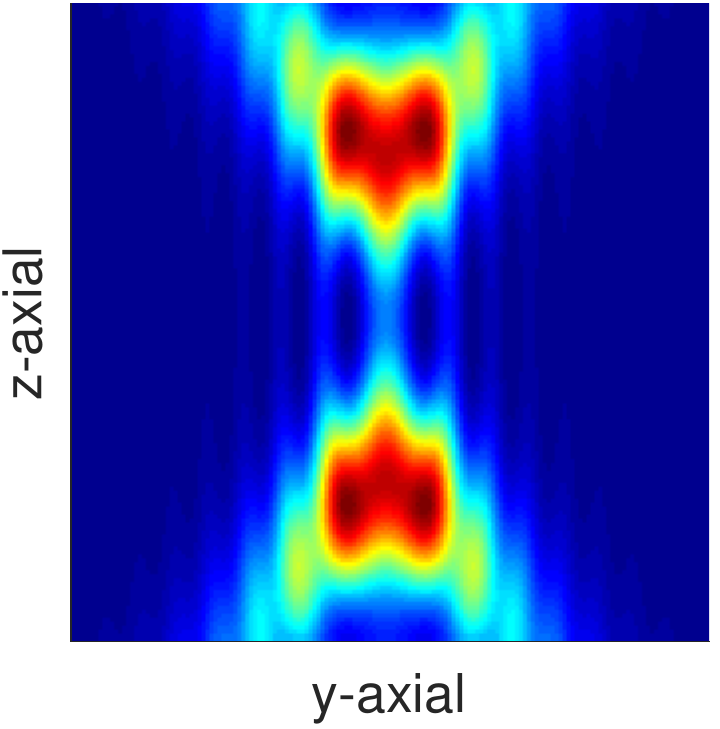}\label{PSF_responses_xCut_4}}
\subfigure[$z=0$]{\includegraphics[height=0.12\textwidth]{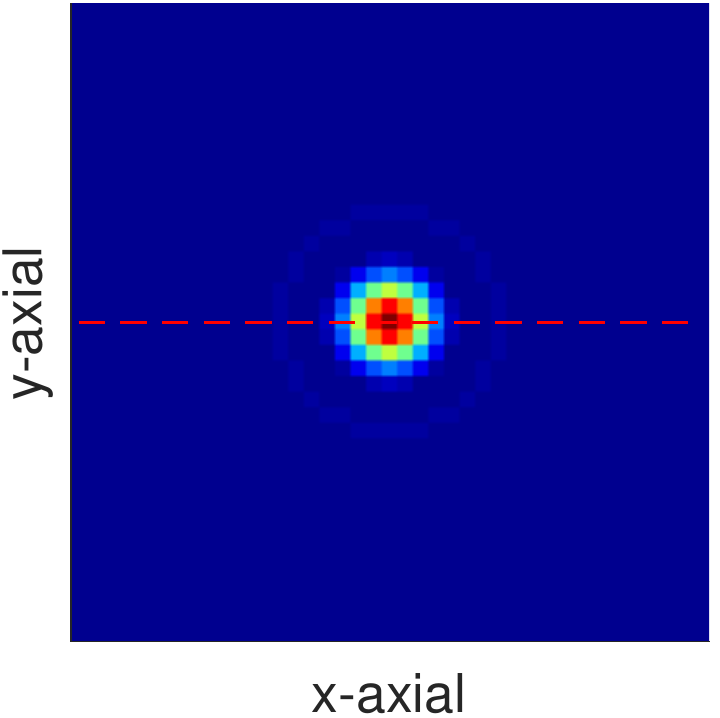}\label{PSF_responses_zCut_0}}
\subfigure[$z=4$]{\includegraphics[height=0.12\textwidth]{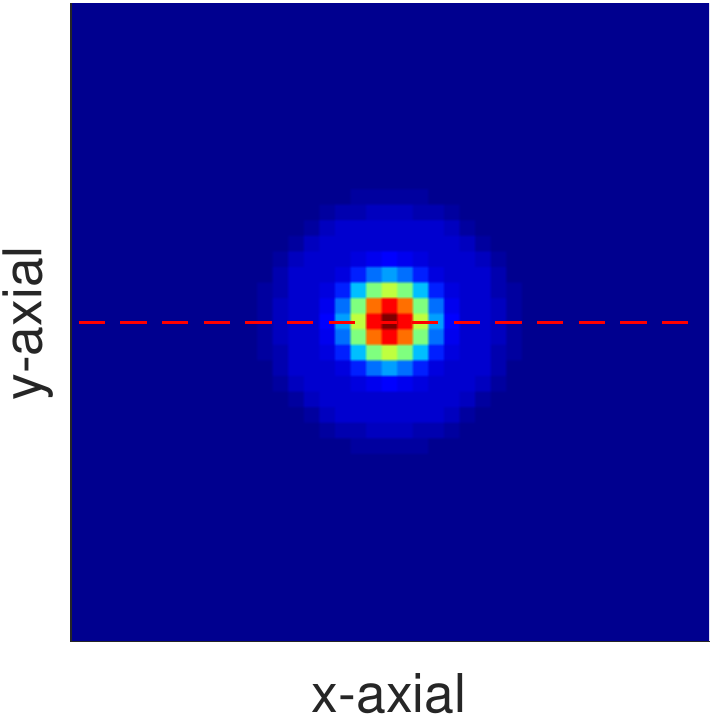}\label{PSF_responses_zCut_4}}
}
\centerline{
\subfigure[$x=0$]{\includegraphics[height=0.12\textwidth]{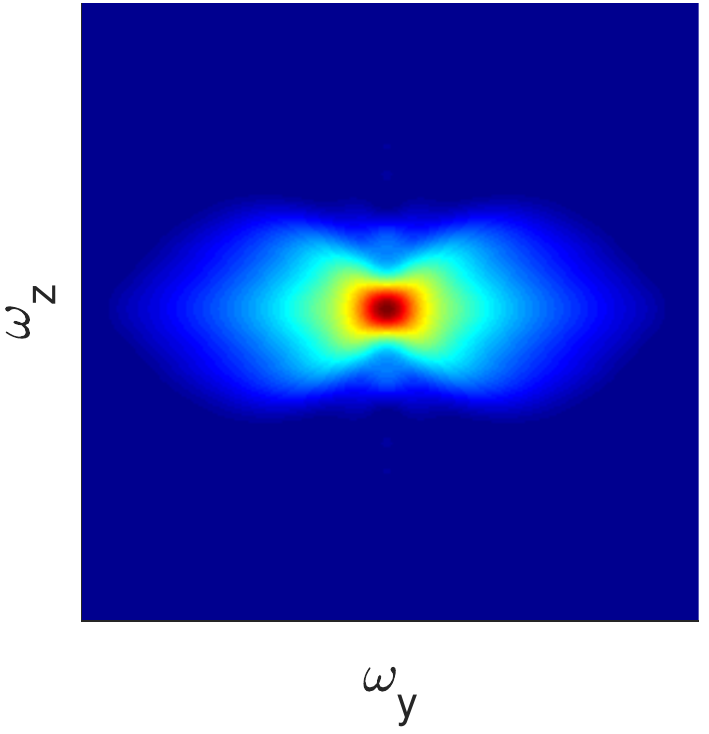}\label{PSF_frequency_xCut_0}}
\subfigure[$x=4$]{\includegraphics[height=0.12\textwidth]{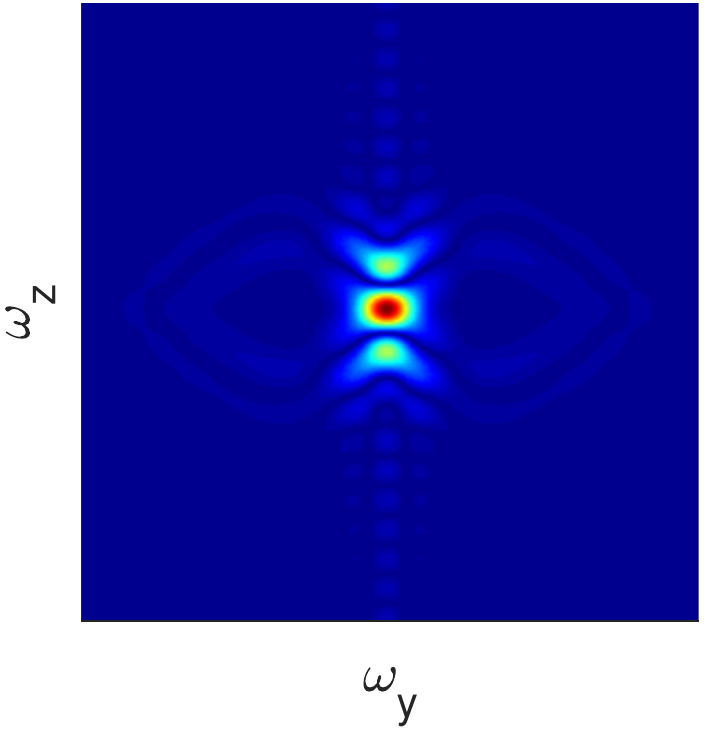}\label{PSF_frequency_xCut_4}}
\subfigure[$z=0$]{\includegraphics[height=0.12\textwidth]{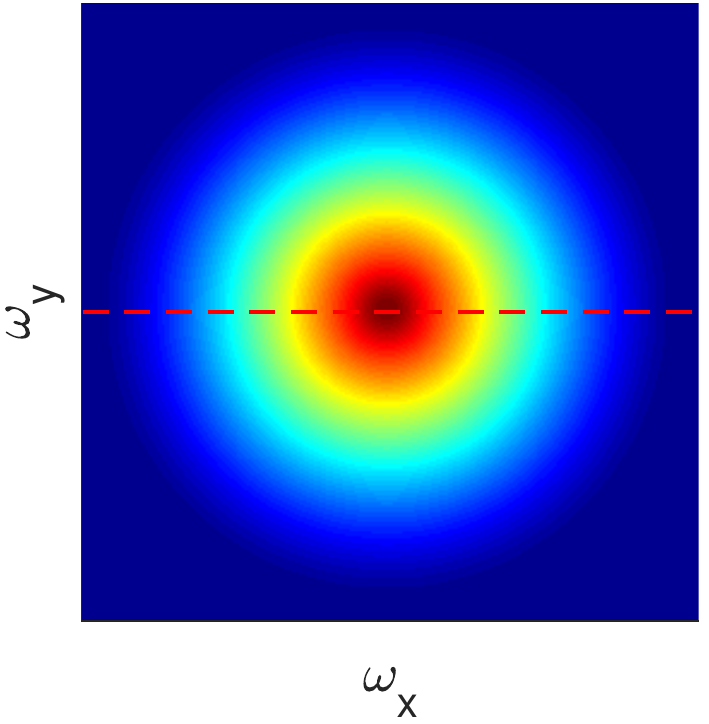}\label{PSF_frequency_zCut_0}}
\subfigure[$z=4$]{\includegraphics[height=0.12\textwidth]{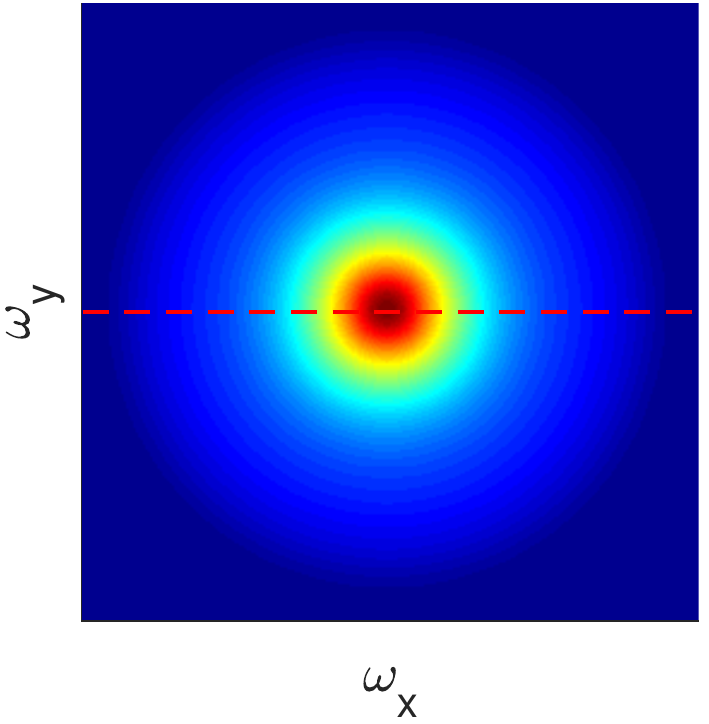}\label{PSF_frequency_zCut_4}}
}
\centering
\caption{2D Plots of the PSF model defined in \ref{equation_PSF_1}, displayed at different x- and z- cut levels shown in both the spatial i.e. first row from (a)-(e) and frequency domains i.e. second row from (e)-(h)}
\label{PSF_responses_xzCuts}
\end{figure}

In Figures \ref{PSF_responses_xCut_0} and \ref{PSF_responses_xCut_4}, we show a mathematical model of the PSF function using Equation \ref{equation_PSF_1} in the axial plane (YZ) at two different $x$-levels. When the objective lens is in focus with respect to lateral plane $X$ i.e. $x=0$, the extent of the frequency domain-PSF is much wider in the $y$-direction than when the $x$-level is out-of-focus $x=4$, shown in Figures \ref{PSF_frequency_xCut_0} and \ref{PSF_frequency_xCut_4}. This shows the importance of proper lateral alignment of the objective lens to mitigate lateral diffraction.

In Figures \ref{PSF_responses_zCut_0} and \ref{PSF_responses_zCut_4}, we show the PSF in the lateral plane (XY) at two different $z$-levels. Note that the lens is in-focus along the optical axis when the $z$-level is zero, causing the model to contain only optical aberration (mainly wavefront) effects. A non-zero $z$-level corresponds to an out-of-focus lens along the optical axis, causing the model to contain both optical aberration and out-of-focus effects. The frequency plots are also shown in Figures \ref{PSF_frequency_zCut_0} and \ref{PSF_frequency_zCut_4}.

\begin{figure}[htp]
\centerline{
\subfigure[$h_{PSF}(x,z)|_{y=0}$]{\includegraphics[height=0.125\textwidth]{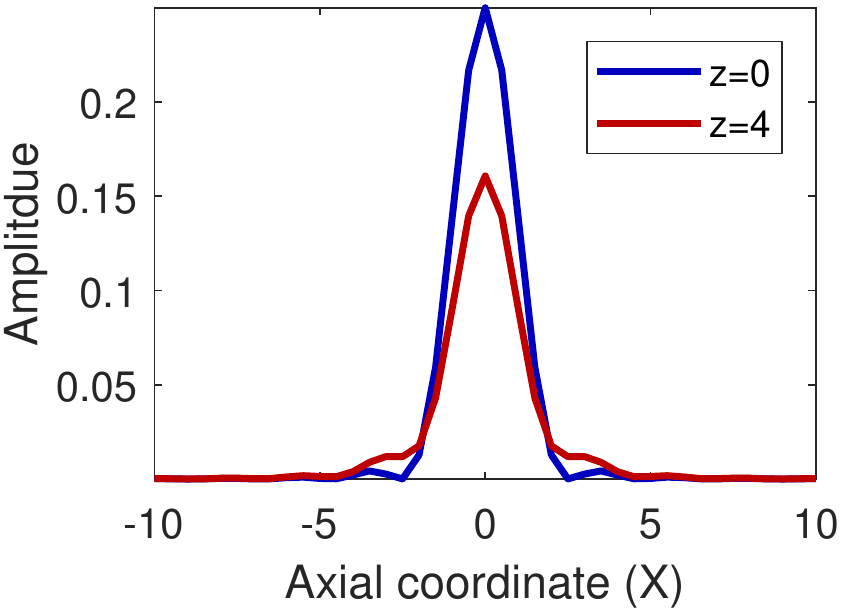}}
\subfigure[$\hat{h}_{PSF}(\omega_x)|_{z,y=0}$]{\includegraphics[height=0.125\textwidth]{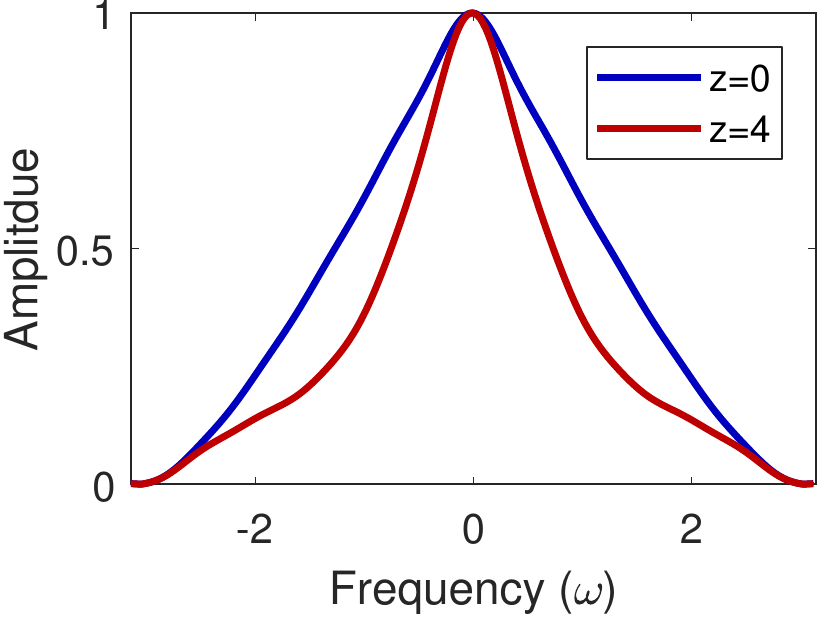}}
}
\centering
\caption{The corresponding 1D PSF models taken along $y=0$ (see the red dashed lines from z- cuts in Figure \ref{PSF_responses_xzCuts}) from the 2D models are: (a) $z\in\{0,4\}$ in spatial domain, (b) $z\in\{0,4\}$ in frequency domain.}
\label{PSF_responses_zCut_yCut}
\end{figure}

The corresponding one-dimensional PSF models are also shown in Figure \ref{PSF_responses_zCut_yCut} where they are taken along $y=0$ (indicated by the dotted red line) from the two-dimensional models shown in Figure \ref{PSF_responses_xzCuts}. Note how the one-dimensional spatial-domain PSF is limited in amplitude for the out-of-focus case (i.e. $z\neq 0$) compared to the in-focus case ($z=0$).

\subsection{HVS-M Kernel Design}
\begin{figure}[htp]
\centerline{
\includegraphics[width=0.3\textwidth]{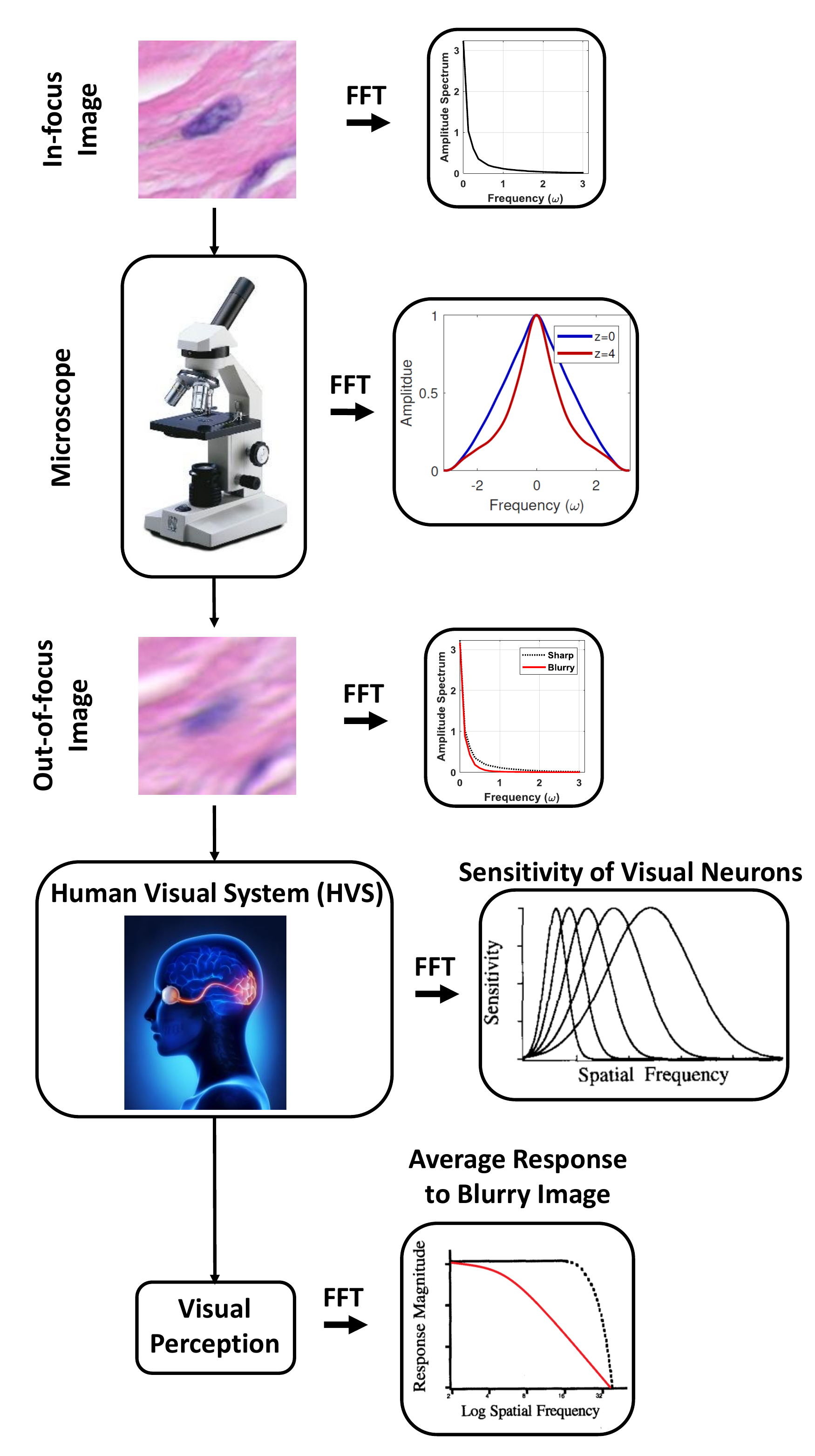}
}
\caption{Out-of-focus blurry microscopy image observed by the human visual system (HVS). The in-focus image of a tissue specimen has a high frequency fall-off characteristic, which is further aggravated by the PSF response of the microscope. The human visual system is known to boost the high-frequency information and perform a deblurring operation, and our synthesized HVS-M kernels are designed to mimic this behavior.}
\label{HVS_modeling}
\end{figure}

In the overall digital pathology process, the PSF described in \ref{equation_PSF_1} i.e. $h_{\text{PSF}}$ can be seen as a low-pass filter (LPF) applied to a natural image, causing its high-frequency components to attenuate and appear blurred. It is also known \cite{field1987relations, brady1995s, field1997visual} that the human visual system (HVS) uses a visual sensitivity response to boost the high frequency response of an image that can be seen as a deblurring operation. See Figure \ref{HVS_modeling} for a diagram of this process.

Inspired by the human visual system's high-frequency-boosting behavior, we propose to synthesize a HVS-like kernel to reverse the high-frequency attenuating behavior in digital pathology images resultant from the PSF of the scanner. We hypothesize that, by boosting the high-frequency information of a scanned digital pathology image that is known to characterize image focality, we will be able to characterize the overall image focus quality.

Here, we design a frequency-equalizing kernel $h_{\text{HVS-M}}$ to mimic the behavior of the human visual system in boosting the high-frequency information attenuated by the modeled PSF:
\begin{align}
	h_{\text{PSF}}(r,z^*) * h_{\text{HVS-M}}(r) = \delta(r),r\in[-r_{max},r_{max}].
	\label{sec_hvs_m_1}
\end{align}
Note that $z^*$ is the optimal microscope objective focal length where by accurate adjustment of this distance we count for both out-of-focus blur and natural image frequency falloff behavior. In particular, we are interested in balancing the frequency falloff in a bounded Fourier transform domain
\begin{align}
\hat{h}_{\text{HVS-M}}(\omega) = \hat{h}_{\text{PSF}}(\omega,z^*)^{-1},\omega\in[-\omega_c,\omega_c].
\label{sec_hvs_m_2}
\end{align}

We model such an enhancement by defining the HVS filter as a linear combination of even-derivative operators
\begin{align}
 h_{\text{HVS-M}}(r,z^*) \equiv \sum_{n=1}^N c_n d_{2n}(r)
\label{sec_hvs_m_3}
\end{align}
where $d_{2n}(r)=d^{2n}/dr^{2n}$ is the $2n$th derivative operator. The Fourier transform of the model in \ref{sec_hvs_m_3} gives
\begin{align}
 \hat{h}_{\text{HVS-M}}(\omega,z^*) = \sum_{n=1}^N c_n \hat{d}_{2n}(\omega) = \sum_{n=1}^N (-1)^n c_n \omega^{2n}.
\label{sec_hvs_m_4}
\end{align}

The unknown coefficients $\{c_n\}_{n=1}^N$ are determined by fitting the response \ref{sec_hvs_m_4} to the inverse of the modeled PSF up to a threshold frequency $\omega_t$
\begin{align}
\underset{\{c_n\}_{n=1}^N}{\mathtt{argmin}}||\hat{h}_{PSF}^{-1}(\omega)-\sum_{n=1}^N c_n (-1)^n\omega^{2n}||_2,\omega\in[0,\omega_t].
\label{sec_hvs_m_5}
\end{align}
This threshold is defined such that to avoid fitting instability on high frequency bands ($\hat{h}_{PSF}^{-1}(\omega)\leq 30$ in our design case). 

Once the optimum coefficients $\{c^*_n\}_{n=1}^N$ are obtained from \ref{sec_hvs_m_5}, we construct the discrete model to the filter design in \ref{sec_hvs_m_4}. We synthesize the even-order derivative filters $\hat{d}_{2n}$ as low-pass filters up to a cutoff frequency $\omega_c$ (in order to balance fitting spurious high-frequency noise and retaining useful high-frequency information) using the MaxPol filter library solution \cite{HosseiniPltaniotis_MaxPol_TIP_2017, HosseiniPltaniotis_MaxPol_SIAM_2017} which is optimized to approximate
\begin{align}
\hat{d}_{2n}(\omega) \approx
\begin{cases}
(-1)^n\omega^{2n},0\leq\omega\leq\omega_c\\
0,\omega\geq\omega_c
\end{cases}.
\end{align}

\begin{figure}[htp]
\centerline{
\subfigure[$\hat{h}_{\text{PSF}}(\omega,z^*)^{-1}$]{\includegraphics[height=0.125\textwidth]{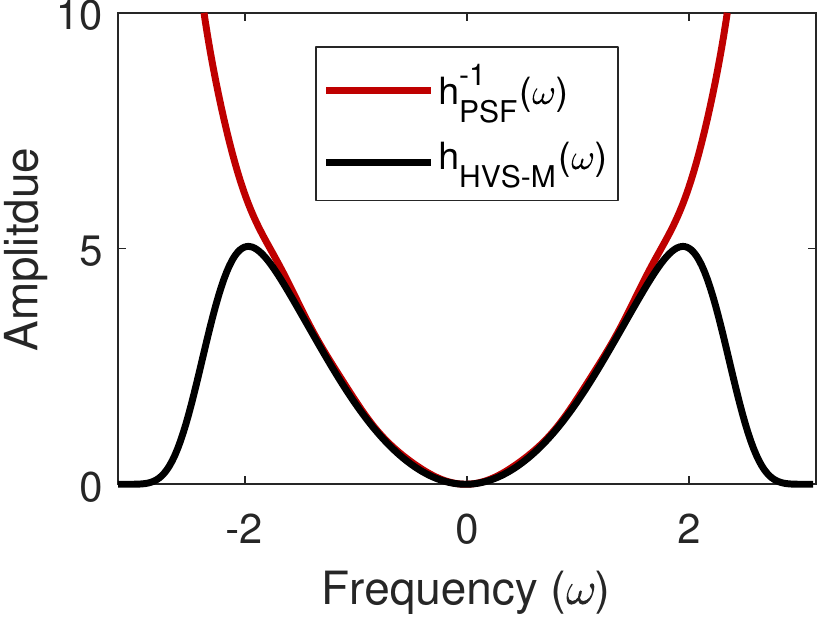}}
\subfigure[$h_{\text{HVS-M}}$]{\includegraphics[height=0.125\textwidth]{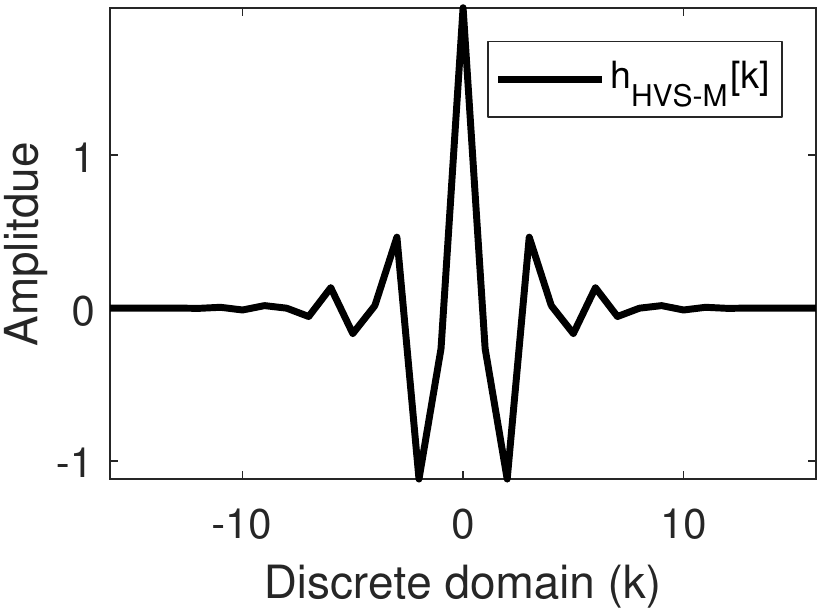}}
}
\centering
\caption{Approximation of HVS-M kernel using MaxPol filter library solution. The optimal coefficients $\{c^*_n\}_{n=1}^N$ are approximated and used to weigh the linear combination of MaxPol even derivative kernels for inverse kernel reconstruction. The inverse spectrum $\hat{h}_{\text{PSF}}(\omega,z^*)^{-1}$ shown above corresponds to the PSF shown in Figure \ref{PSF_responses_zCut_yCut} for $z=4$.}
\label{PSF_responses_zCut_yCut_MaxPol_approximate}
\end{figure}

Figure \ref{PSF_responses_zCut_yCut_MaxPol_approximate} demonstrates an example of an approximated HVS-M kernel. The inverse profile $\hat{h}_{\text{PSF}}(\omega,z^*)^{-1}$ is first fitted by the frequency polynomials defined in (\ref{sec_hvs_m_5}), and the optimal coefficients $\{c^*_n\}_{n=1}^7$ are attained. These coefficients are used to combine different MaxPol even-derivative filters to construct the HVS-M kernel. All of the MaxPol filters are regulated by a cutoff frequency $\omega_c=2$ for discrete approximation.

\section{Patch-Based Level Focus Quality Assessment} \label{sec:focus}
The constructed HVS-M kernel defined in the previous section can now be applied as a convolution operation to partly modify the frequency amplitudes of an out-of-focus image in order to extract sharpness features for focus quality grading in WSI image patch. The HVS-M kernel provides a meaningful transformation where the edge features across a wide frequency spectrum are balanced and become comparable with each other, and are known to be related to focus quality. Furthermore, by assigning a single focus quality score for each patch enables localized focus quality heatmap generation described in the next section.

\begin{algorithm}
	\caption{Proposed FQPath scoring for WSI patches}
	\label{algo_score}
	\KwData{Digital pathology image patch $I\in\mathbb{R}^{N_1\times N_2\times 3}$, 
	
	Out-of-Focus Microscopy PSF $h_{PSF}(z^*)\in\mathbb{R}^{2r_{max}+1}$,
	
	optimal parameters $\{z^*, \omega_c^*, m^*\}$}
	\KwResult{Out-of-focus score $s\in\mathbb{R}$}
	Synthesize inverse response of out-of-focus kernel
	$\{c_n^*\}_{n=1}^N\gets\underset{c_n}{\mathtt{argmin}}||\hat{h}_{\text{PSF}}(\omega)^{-1}-\sum\limits_{n=1}^N c_n (-1)^n\omega^{2n}||_2$ for $\omega\in[0,\omega_t]$
	
	Construct HVS-M using MaxPol lowpass derivatives $h_{\text{HVS-M}}[k]=\sum\limits^{N}_{n=1}{c_n^* d_{2n}[k]}$
	
	Decompose grayscale image features using HVS-M $(F_x,F_y)= I*(h_{\text{HVS-M}}^T,h_{\text{HVS-M}})$
	
	Activate features $F^R_x=\max(F_x, 0), ~F^R_x=\max(F_y, 0)$
	
	Vectorize only positive features $v\equiv\mathtt{vec}({F^R_x}^{+}, {F^R_y}^{+})$
	
	Find $95$th percentile of cumulative distribution function of the feature vector $\sigma_{0.95}=\mathtt{CDF}(v,95\%)$

	Find the proportion of features to be retained by 
	$P= 0.25(1-\mathtt{tanh}(60(\sigma_{0.95}-0.095)))+0.09$
		
	Find the $\#$pixels to be retained $N_{\text{pix}}=P N_1 N_2$
	
	Find the $\ell_{{1}/{2}}$-norm feature vectors $
	F=||({F^R_x}^{+},{F^R_y}^{+})||_\frac{1}{2}$
	
	Retain $\overline{F} = \textit{sort}_{\text{descend}}(F)_k,~k\in\{1,\hdots,N_{\text{pix}}\}$
	
	Calculate $\mu_{m^*}=\mathbb{E}[(\overline{F}-\mu_0)^{m^*}],~\text{where}~\mu_0=\mathbb{E}[\overline{F}]$
	
	Record the score value by $s=-\log\mu_{m^*}$
\end{algorithm}

\begin{figure}[htp]
	\centerline{
	\subfigure[Original Image]{\includegraphics[height=0.12\textwidth]{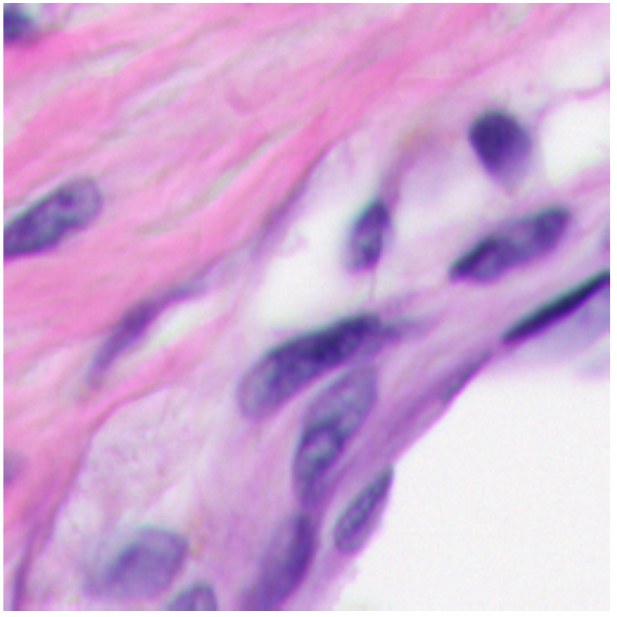}}
	\subfigure[$F$]{\includegraphics[height=0.12\textwidth]{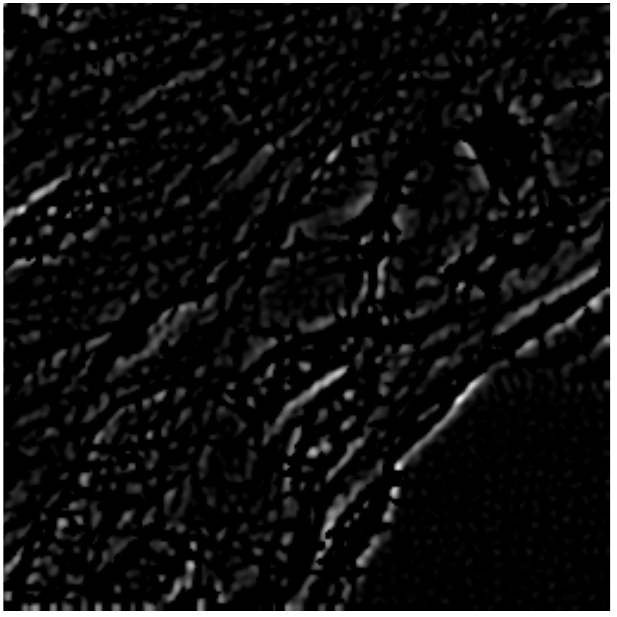}\label{FQM_decomposed_images_F}}
	\subfigure[Sorted features $\bar{F}$]{\includegraphics[height=0.125\textwidth]{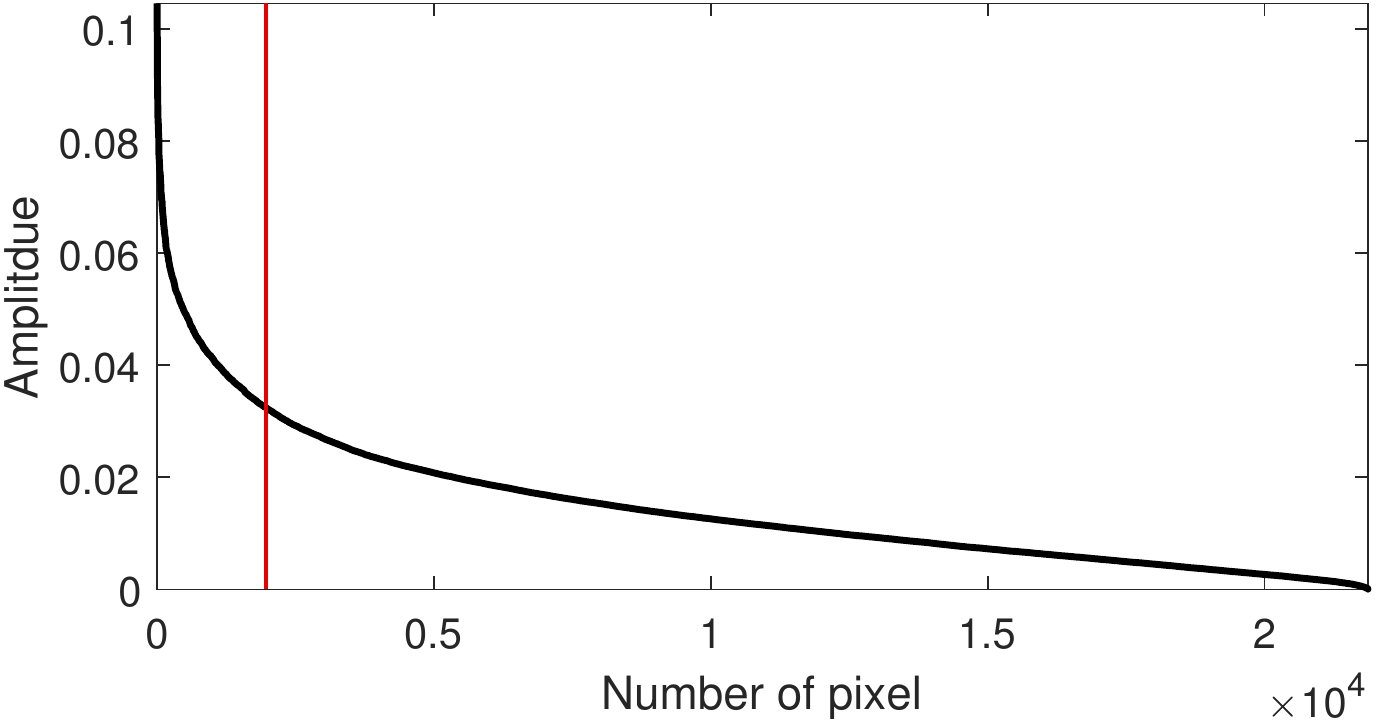}\label{FQM_decomposed_images_vector}}
	}
	\centerline{
	\hspace{.025in}
	\subfigure[$F^R_x$]{\includegraphics[height=0.12\textwidth]{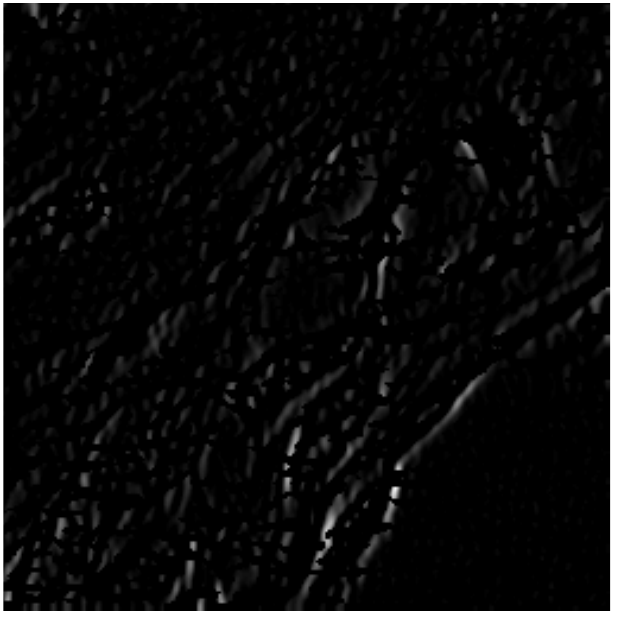}\label{FQM_decomposed_images_h_p}}
	\subfigure[$F^R_y$]{\includegraphics[height=0.12\textwidth]{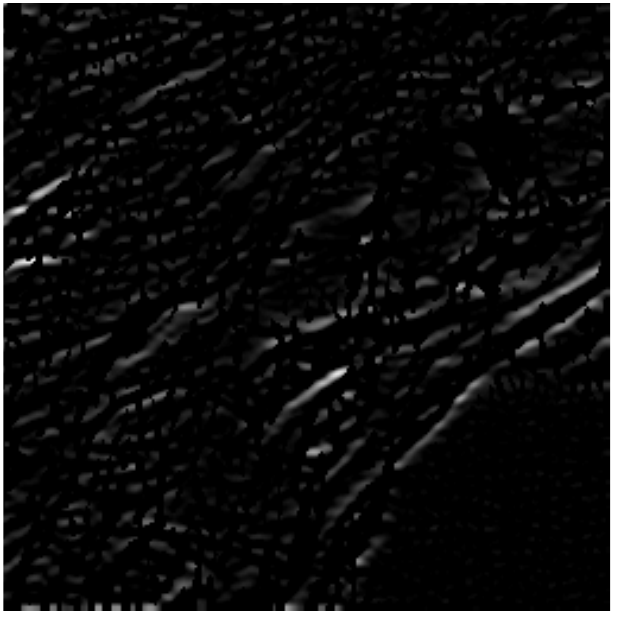}\label{FQM_decomposed_images_v_p}}
	\subfigure[$F_x$]{\includegraphics[height=0.12\textwidth]{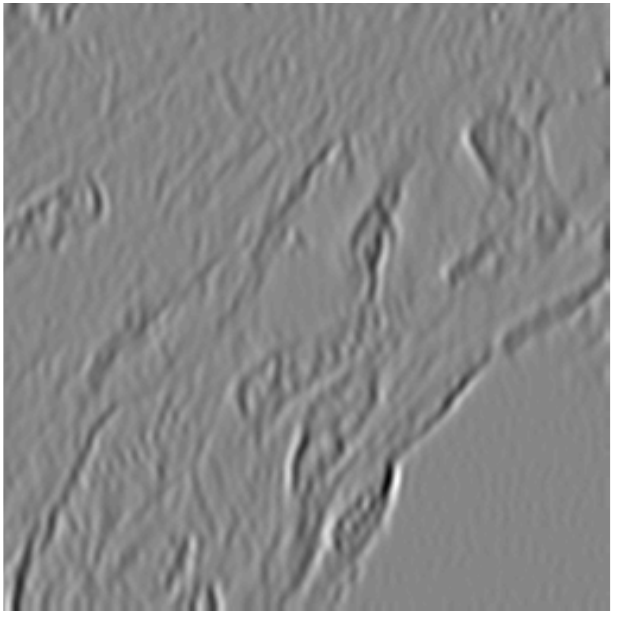}\label{FQM_decomposed_images_h}}
	\subfigure[$F_y$]{\includegraphics[height=0.12\textwidth]{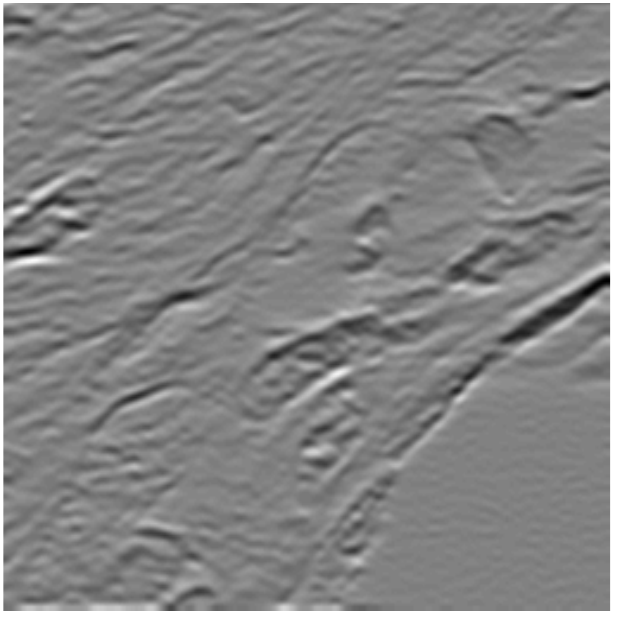}\label{FQM_decomposed_images_v}}
	}
	\centering
	\caption{Demo images after each single operation for metric analysis defined in Algorithm \ref{algo_score}. The input image is decomposed by HVS-M kernel (d) horizontally and (e) vertically. The positive features are activated from both decomposed images shown in (f) and (g), where the superimposed feature map is shown in (b). The features are sorted in (c) and an adaptive threshold is applied to keep $N_{\text{pix}}$ largest feature values.}
	\label{FQM_decomposed_images}
\end{figure}

Our metric requires the input image patch to be a grayscale image sized $1024\times 1024$. The focus score procedure is defined in Algorithm \ref{algo_score} and a select step-by-step visualization of the procedure is shown in Figure \ref{FQM_decomposed_images}. The algorithm requires three parameters $z^*$, $\omega_c^*$, $m^*$ to be tuned as follows:
\begin{itemize}
	\item $z$-axial PSF coordinate $z^*$ used for the PSF model $h_{\text{PSF}}(z^*)$ defined in (\ref{equation_PSF_1})
	\item Cutoff frequency $\omega_c^*$ used for setting the cutoff response of the MaxPol lowpass derivative basis filters
	\item Central moment $m^*$ used for quantifying the $m^*$th order energy of the focus feature vector
\end{itemize}
To tune these parameters, we performed a grid search to yield the best correlation accuracy between the subjective (ground-truth $z$-levels) and objective quality scores in our tuning database, which was a subset of full-FocusPath database introduced in \cite{hosseini2018focus}.

In our proposed focus quality scoring in algorithm \ref{algo_score}, first, we fit the inverse PSF in the frequency domain with frequency polynomials up to the $N$th order to obtain the optimized coefficients $\{c^*_n\}^N_{n=1}$. Then, the coefficients are used to weight the linear combination of one-dimensional even-derivative MaxPol filters for synthesizing the HVS-M filter (this filter is calculated once and fixed for subsequent image scoring). We then separate the input image with the HVS-M filter along the horizontal and vertical axes (see Figures \ref{FQM_decomposed_images_h} and \ref{FQM_decomposed_images_v}) and exclude negative features with the ReLU function to avoid blurring adjacent features with redundant information (see Figures \ref{FQM_decomposed_images_h_p} and \ref{FQM_decomposed_images_v_p}). The feature map is vectorized and an adaptive proportion of the largest values are retained, using a nonlinear function, based on the $95$th percentile of the feature vector's probability distribution function (PDF). This ensures that only the most dominant focus-relevant features are preserved. The energy of the horizontal and vertical feature is taken in $\ell_{1/2}$-norm space to promote feature sparsity along the horizontal and vertical directions (see Figure \ref{FQM_decomposed_images_F}). Finally, we calculate the $m^*$th central moment of the most dominant values from the feature vector (see Figure \ref{FQM_decomposed_images_vector}) and take its negative logarithm to obtain the final focus quality score. Note that the focus quality score is inversely proportional to the focus quality, i.e. a lower score indicates better focus quality.

\section{WSI-Based Level Focus Quality Assessment}\label{sec:heatmap}
Previously, we explained the focus quality measure at the WSI patch level, but for this measurement to be useful in a routine digital pathology workflow, it needs to be conducted at the tissue slide level. This way, the focus quality measure can be used to generate a local focus quality map (known as a heatmap) for an entire scanned slide. For a given whole slide image, we propose to divide it into patches of size $1024\times 1024$ px, calculate the focus quality score for each patch separately, apply an inverse Gaussian projection (to linearize the focus quality gradation), and combine the patch scores into a slide-level focus quality heatmap. In this section, we describe the WSI database that we used for testing, our proposed inverse Gaussian projection linearization technique, and the subjective whole-slide scoring we conducted to validate our generated heatmaps.

\subsection{Inverse Gaussian Projection}
\begin{figure}[htp]
	\centerline{
  \subfigure[Train prof.]{\includegraphics[height=0.125\textwidth]{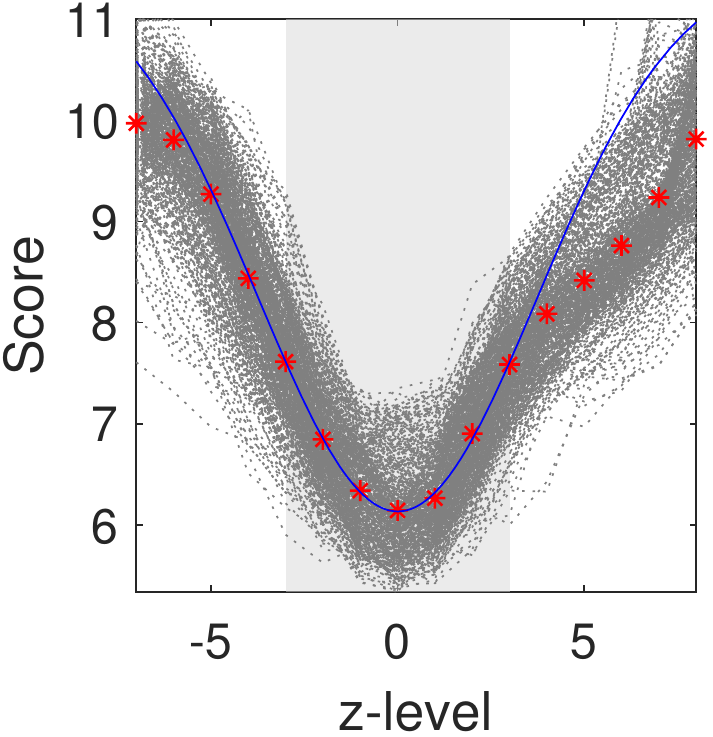}\label{reg_train_profiles}}
	\subfigure[Proj. train prof.]{\includegraphics[height=0.125\textwidth]{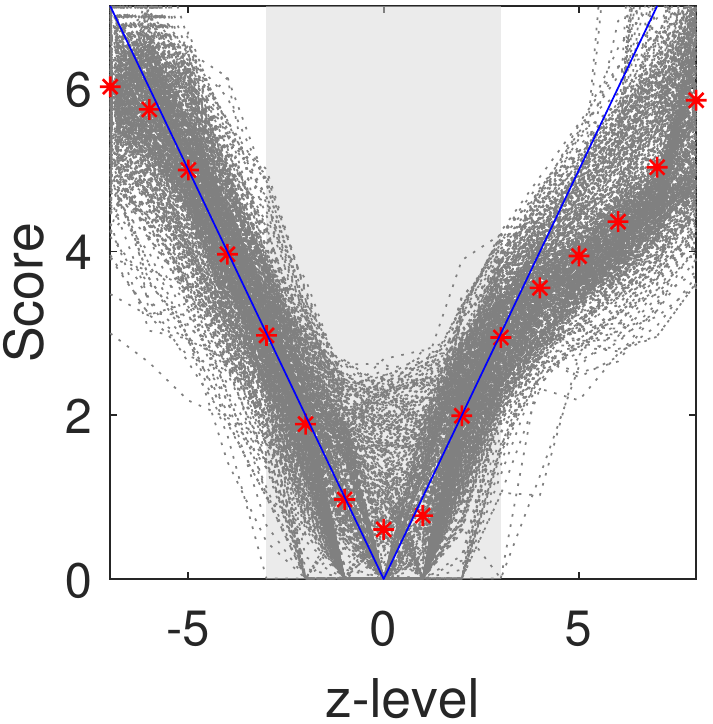}\label{proj_train_profiles}}
	\subfigure[Test prof.]{\includegraphics[height=0.125\textwidth]{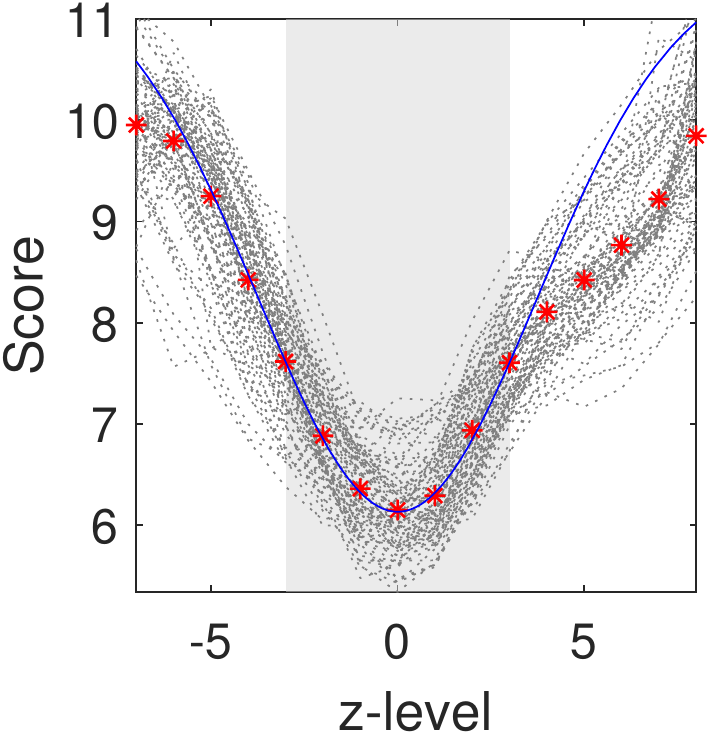}\label{reg_test_profiles}}
	\subfigure[Proj. test prof.]{\includegraphics[height=0.125\textwidth]{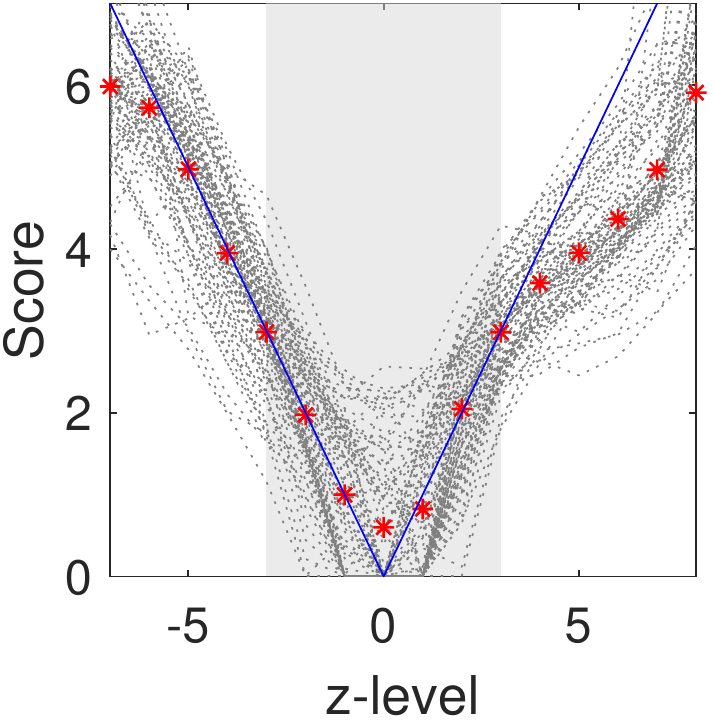}\label{proj_test_profiles}}
}
\centering
\caption{Plots of $z$-profile scores, showing the individual profiles as grey dotted lines, the mean profiles as red asterisks, and the linear/Gaussian fits as blue solid lines, with the central grey region representing the fitting window: (a) Regular training $z$-profile scores, (b) Projected training $z$-profile scores, (c) Regular testing $z$-profile scores, (d) Projected testing $z$-profile scores}
	\label{profiles}
\end{figure}

One additional step is used to prepare the patch-based focus quality measure discussed in the previous section for slide-level heatmap generation - inverse Gaussian projection. This ensures that the score-$z$ profile (the relationship between the focus quality score and the known ground-truth $z$-level) is accurately linearized, which is important for proper visual perception of focus quality gradations and local extrema, since a non-linearized scale causes the heatmap to saturate near high and low focus quality values. We selected the inverse Gaussian projection to be the linearizing operation because we noticed that the score-$z$ profiles tended to form an inverted Gaussian-like bell shape (as seen in Figures \ref{reg_train_profiles} and \ref{reg_test_profiles}).

To implement and validate the inverse Gaussian projection, we take the full, unreleased version of the FocusPath database \cite{hosseini2018focus} (540 lateral profiles$\times$16 $z$-levels$=$8640 images in total) to be the training set for calibrating the projection, and a separate set of digital pathology images to be the test set (90 lateral profiles$\times$16 $z$-levels$=$1440 images in total) for validating the calibration. All profiles have ground-truth $z$-levels attached to them (integer values ranging from -7 to 8, inclusive). To train (or calibrate) the inverse Gaussian projection, we obtained the regular focus quality scores for these 540 training profiles, plotted them at their respective ground-truth $z$-levels, and found the mean focus quality score for each $z$-level (see the grey dotted lines and red asterisks respectively in Figure \ref{reg_train_profiles}). Then, we fitted a Gaussian 
$f(z) = ae^{-(z-b)^2/c}$ to the inverse mean focus quality score in the middle $z$-level values (i.e. the central grey region $z\in[-3,3]$ in Figure \ref{reg_train_profiles}) to prioritize fitting in the high focus quality region (the optimal values were determined to be $a^*=5.389,b^*=0.005248,c^*=5.301$). For testing, we first inverted the regular focus quality score, passed it through a threshold to ensure a non-negative argument to square root, and then applied the inverse of the fitted Gaussian to the score. See algorithm \ref{algo_projection} for details on the training and testing phases of the projection.

\begin{algorithm}
	\caption{Inverse Gaussian projection on FQPath scores}
	\KwData{Training set of FQPath scores $s_\text{t}(n,z)\in\mathbb{R}$, where $n=\{1,\cdots,540\}$ and $z=\{1,\cdots,16\}$\\
					Single test FQPath score $s\in\mathbb{R}$}
	\KwResult{Projected testing FQPath score $s_\text{P}\in\mathbb{R}$}
	
	Find the mean train set profile $\bar{s}_\text{t}(z)=\mathbb{E}_n[s_\text{t}(n,z)]$
	
	Calculate inverse mean profile $\bar{s}_\text{t\_inv}(z)=\mathtt{max}(\bar{s}_\text{t})-\bar{s}_\text{t}(z)$
	
	Fit Gaussian model to inverse mean profile $\{a^*,b^*,c^*\}\gets\underset{\{a,b,c\}}{\mathtt{argmin}}||\bar{s}_\text{t\_inv}(z)-ae^{-(\frac{z-b}{c})^2}||_2$, where $z\in\{-3,-2,\hdots,2,3\}$
	
	Invert and threshold test set score
	$s_\text{inv}=\mathtt{min}(\mathtt{max}(\bar{s}_\text{t})-s,a^*)$
	
	Map test score by inverse Gauss. $s_\text{P}=c^*\sqrt{-\log{\frac{s_\text{inv}}{a^*}}}+b^*$
	
	\label{algo_projection}
\end{algorithm}

\begin{figure}[htp]
	\centerline{
		\subfigure[Before projection]{\includegraphics[height=0.2\textwidth]{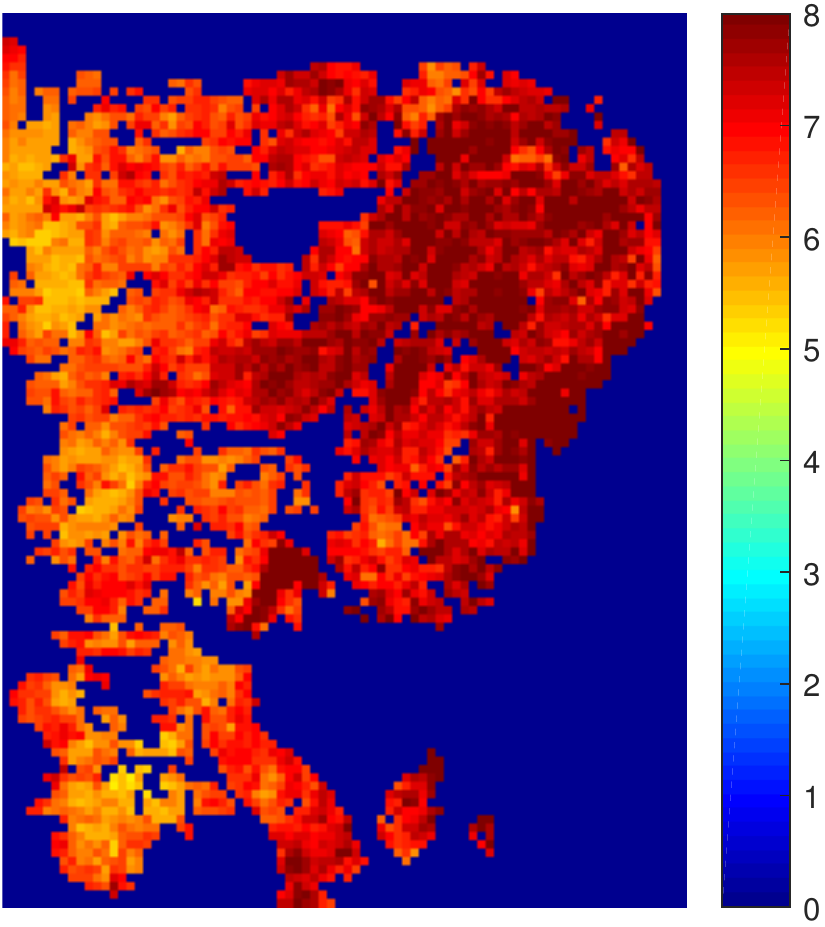}}\hspace{.1in}
		\subfigure[After projection]{\includegraphics[height=0.2\textwidth]{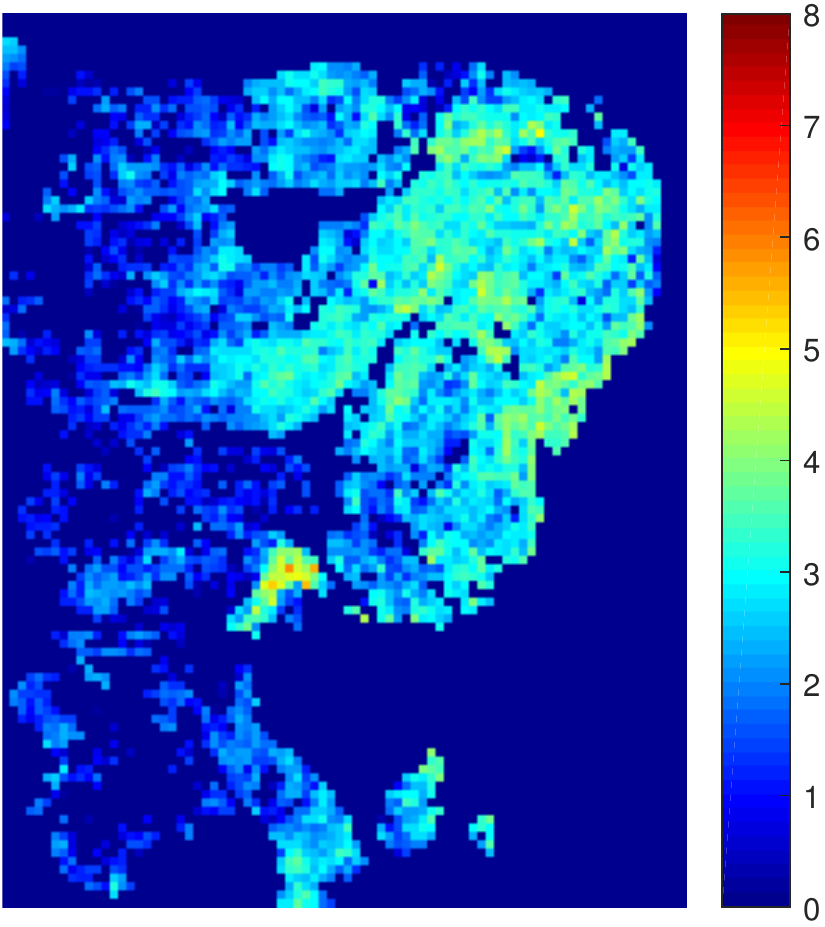}}
	}
	\centering
	\caption{Sample heatmap before and after projection. Note the greater visual saliency of low focus quality regions in the heatmap after projection.}
	\label{heatmap_projection_demo}
\end{figure}

As can be seen in Figure \ref{profiles}, the score-$z$ profiles are significantly more linear near inside the grey fitting window after projection (Figures \ref{proj_train_profiles} and \ref{proj_test_profiles}) than before projection (Figures \ref{reg_train_profiles} and \ref{reg_test_profiles}). Note that, in both the training and testing after-projection profiles, a minority of individual profiles undesirably stagnate near zero within the range $z\in[-2,2]$. This is caused by the threshold operation required for the inverse Gaussian. When applied to an entire slide for heatmap generation, as shown in Figure \ref{heatmap_projection_demo}, it is clear that the inverse Gaussian projection linearizes the heatmap colour scale, enabling better visual perception of focus quality gradations and local extrema.

\subsection{200-WSI Scan Database}
For the purpose of focus quality assessment at the WSI level, we have collected a slide database consisting of 200 different WSIs scanned at 40X magnification ($0.25\mu$m/pixel resolution) without compression using a TissueScope LE1.2 scanner. This database is selected from a larger set of 500 anonymized H\&E tissue glass slides (sized 1''$\times$ $3$'', 1.0 mm in thickness) mainly provided by Southlake Regional Hospital and made available to the authors courtesy of Huron Digital Pathology. The slides were specifically selected to cover the entire histopathological colour range and to exclude slides with excessive preparation imperfections, such as air bubbles and tissue folding. The tissue sections vary in thickness, organ of origin (e.g. brain, kidney, breast, liver, heart), and diagnosis (e.g. healthy, tuberculosis, cancer).

\subsection{Subjective Whole-Slide Scoring}
\begin{figure}[htp]
    \centerline{
        \includegraphics[height=0.25\textwidth]{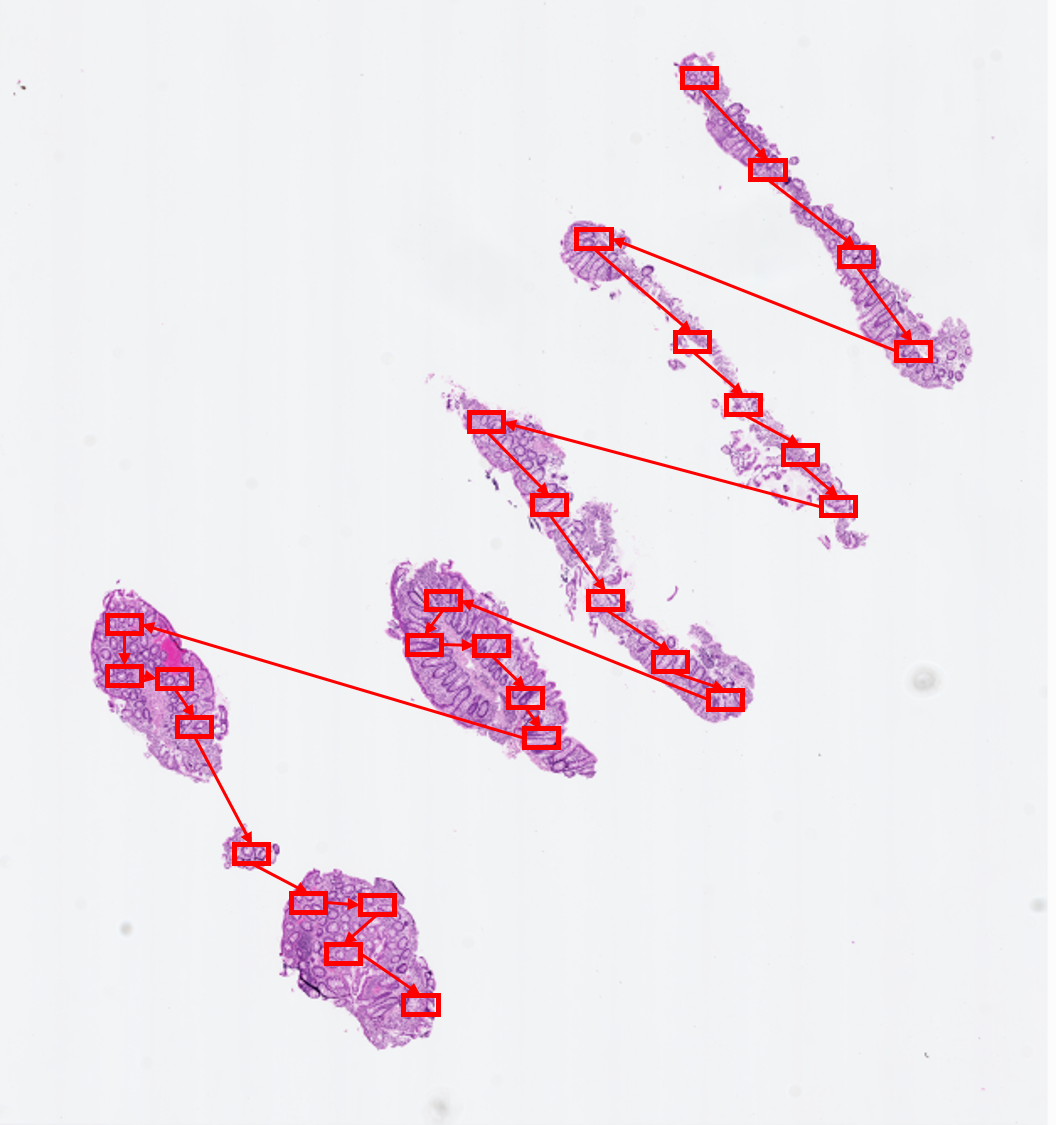}
        \includegraphics[height=0.25\textwidth]{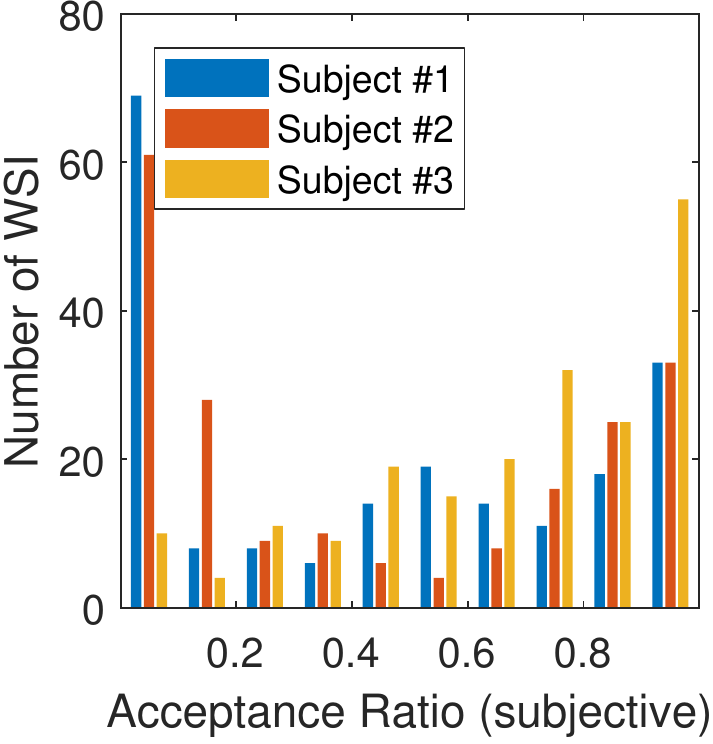}
    }
    \centering
	\caption{Left plot: Sample pre-planned navigation pattern of assessment points through a slide. The assessor navigates through different positions within the tissue thumbnail at $100\%$ image resolution and records each as pass/fail. The ratio of pass points to total points determines the acceptance ratio of WSI sharpness. Right plot: Statistical distribution of acceptance ratio across 200 WSI database and three different subjective ratings.}
	\label{nav_pattern}
\end{figure}

In order to validate the accuracy of the generated focus quality heatmaps, comparison must be made to a subjective score. Hence, we trained three human focus quality assessors to score the subjective quality of each slide at WSI-based level and compare these scores with the generated heatmaps. The assessors were trained on ground-truth views of in-focus and out-of-focus WSI regions displayed at 100\% pixel resolution. A region was labeled “focus-rejected” when more than half of the screen view area was out-of-focus. Accordingly, the following procedure was followed:

\begin{enumerate}
	\item Open the WSI in the HuronViewer program \footnote{HuronViewer is freely available at \url{http://www.hurondigitalpathology.com/resources/}}
	\item Increase the zoom level to 100\% pixel resolution
	\item Using the thumbnail view, mentally plan the slide navigation pattern of assessment points to cover as much of the tissue regions and as evenly as possible (for example see Figure \ref{nav_pattern} (left plot))
	\item Start a one-minute countdown within view of the slide
	\item Using the thumbnail view, navigate the field of view to the first assessment point
	\item Determine the field-of-view (FOV) around the first assessment point to be in-focus (pass) or out-of-focus (fail)
	\item Increment the count of passing and failing focus quality assessment points accordingly
	\item Proceed to the next assessment point (according to the pre-planned slide navigation pattern) and repeat steps 6 to 8 until either one minute has passed or the pre-planned slide navigation pattern has been completed early
\end{enumerate}

As a result, each of the 200 slides was assigned two numbers (pass and fail counts) by the three assessors and an acceptance ratio (number of pass points divided by total points) was calculated from these counts. The statistical distribution of the slide acceptance ratios in the 200-WSI database is shown in Figure \ref{nav_pattern} (right plot). It shows the assessor subjective scores differ due to their varying expertise and quality criteria. One can obtain the mean opinion score (as is standard practice in image quality assessment \cite{wang2004image, sheikh2006statistical}) to create the relative subjective scoring of the whole slide image.

\section{Experiment No.1: Patch-Level Analysis} \label{sec:experiment_1}
In this section, we describe the design, implementation and execution of several experiments at the WSI patch level to evaluate our proposed FQ metric in terms of accuracy, processing speed, and computational complexity. We compare our metric against 10 state-of-the-art NR-FQA metrics, including $\text{S}_{3}$ \cite{vu2012bf}, MLV \cite{bahrami2014fast}, $\text{ARISM}_{\text{C}}$ \cite{gu2015no}, GPC \cite{leclaire2015no}, SPARISH \cite{li2016image}, RISE \cite{li2017no}, MaxPol \cite{mahdi2018image}, HVS-1 and HVS-2 \cite{hosseini2018focus}, and MIFQ \cite{yang2018assessing}.

\subsection{Degree of Accuracy Correlation}
\begin{figure}[htp]
\scriptsize
\centerline{
\subfigure[$\text{S}_3$ \cite{vu2012bf}]{\includegraphics[height=0.12\textwidth]{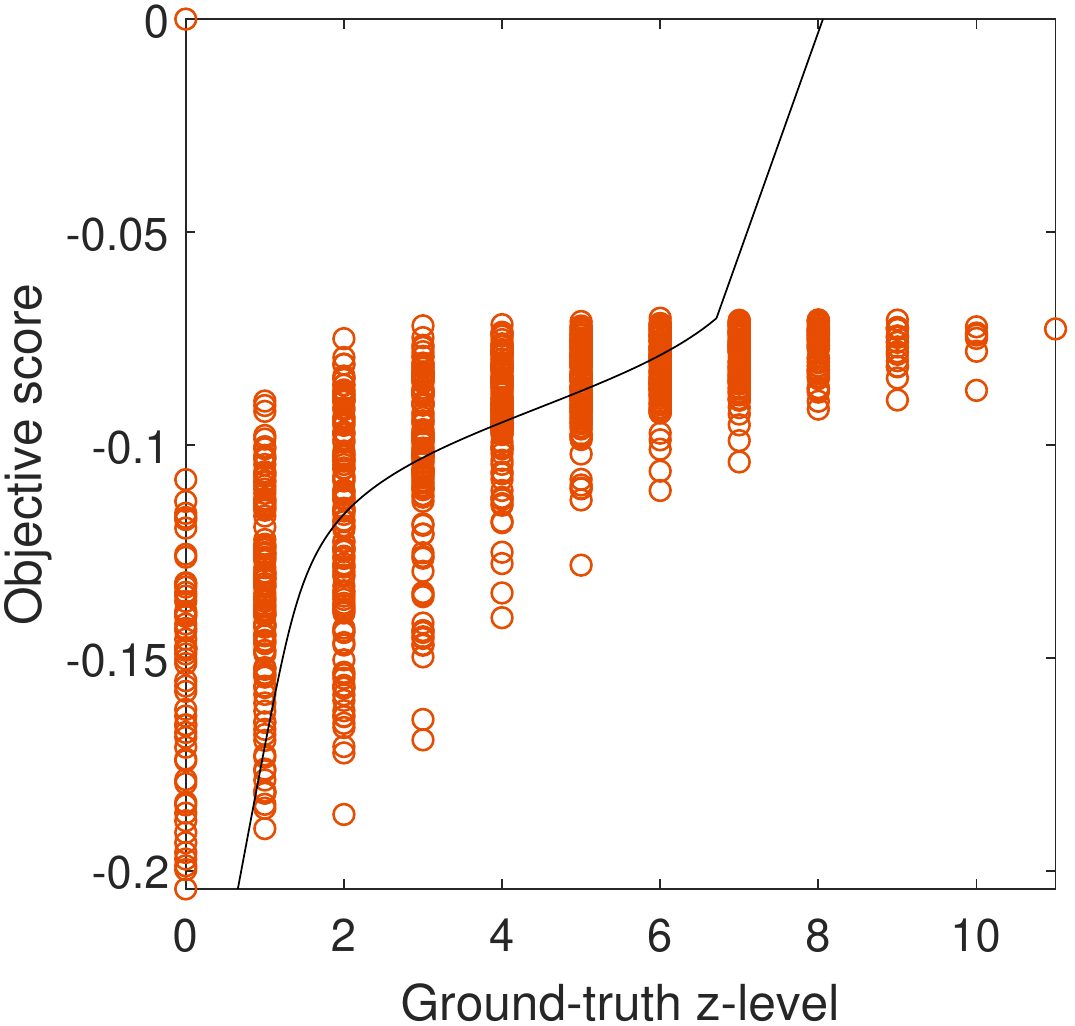}}
\subfigure[MLV \cite{bahrami2014fast}]{\includegraphics[height=0.12\textwidth]{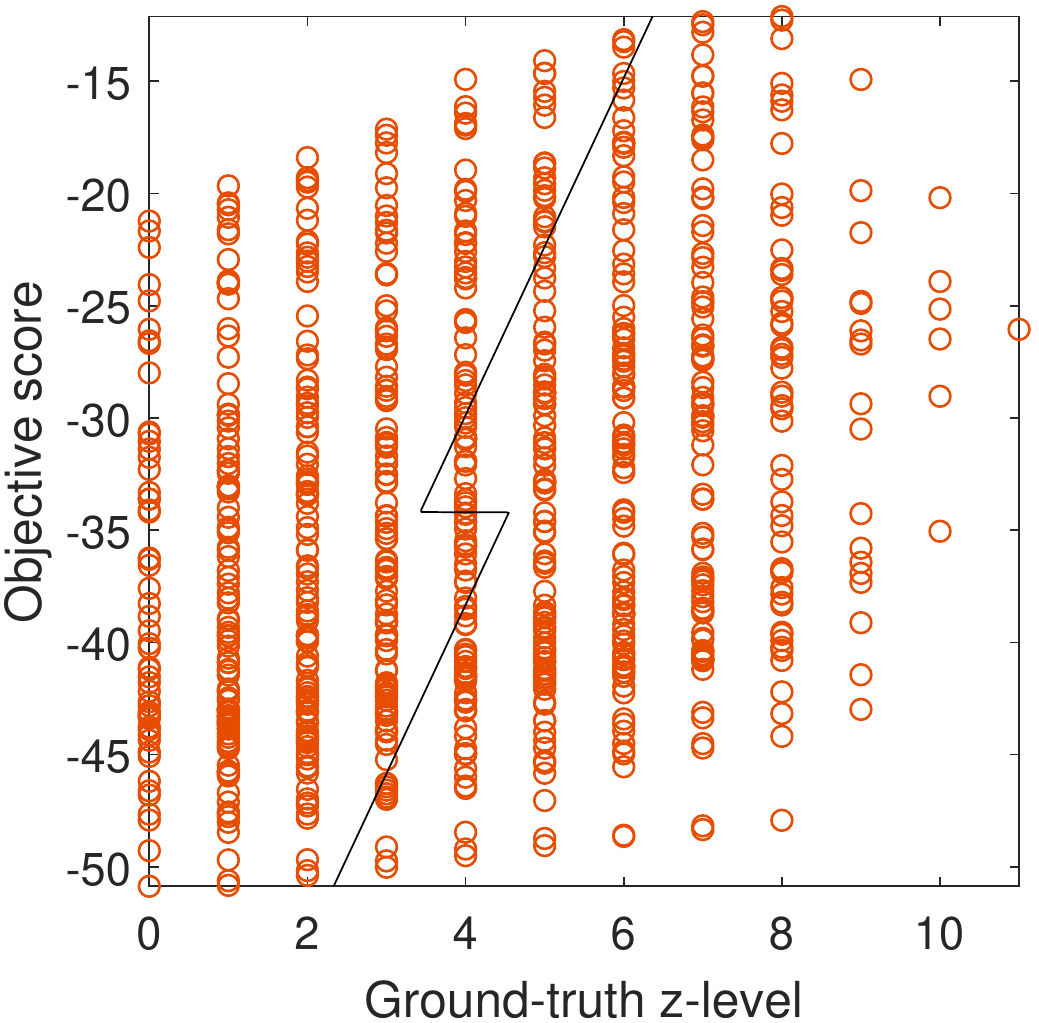}}
\subfigure[$\text{ARISM}_{\text{C}}$ \cite{gu2015no}]{\includegraphics[height=0.12\textwidth]{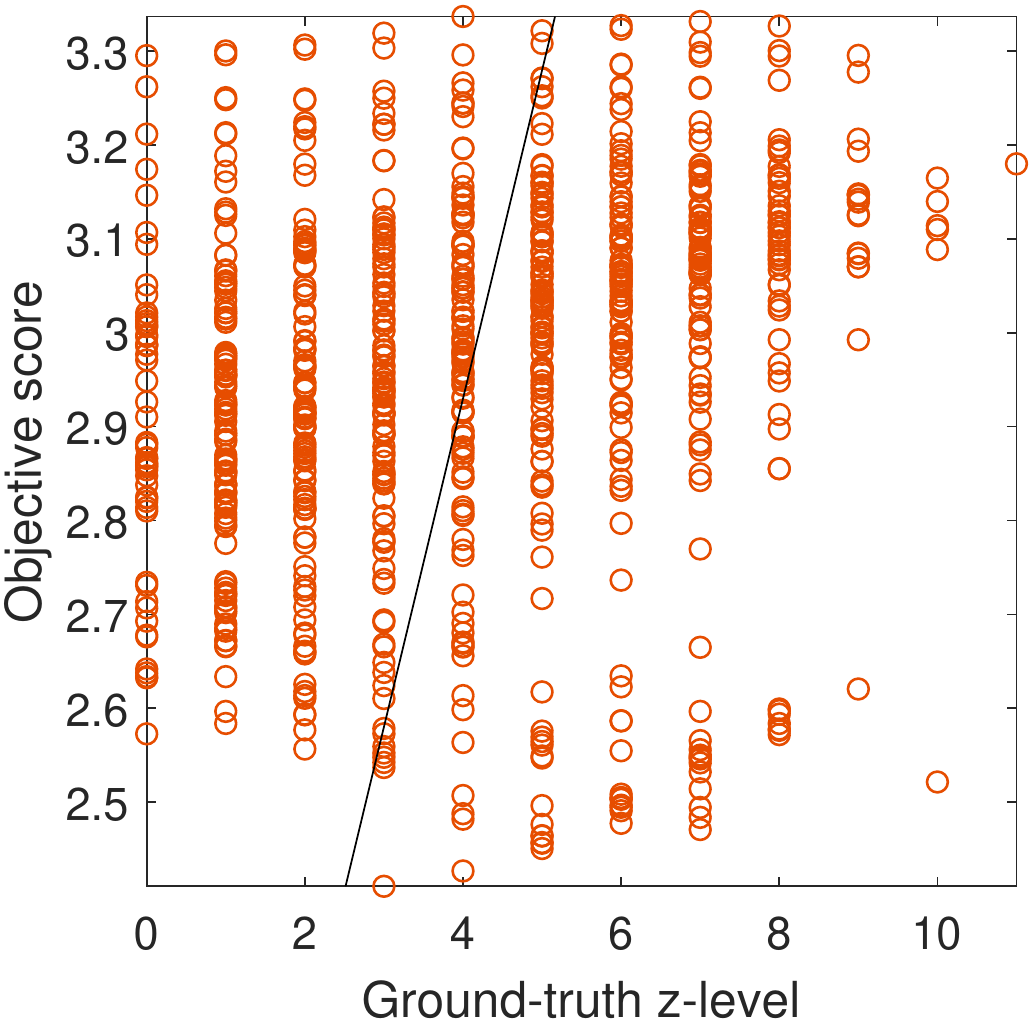}}
\subfigure[GPC \cite{leclaire2015no}]{\includegraphics[height=0.12\textwidth]{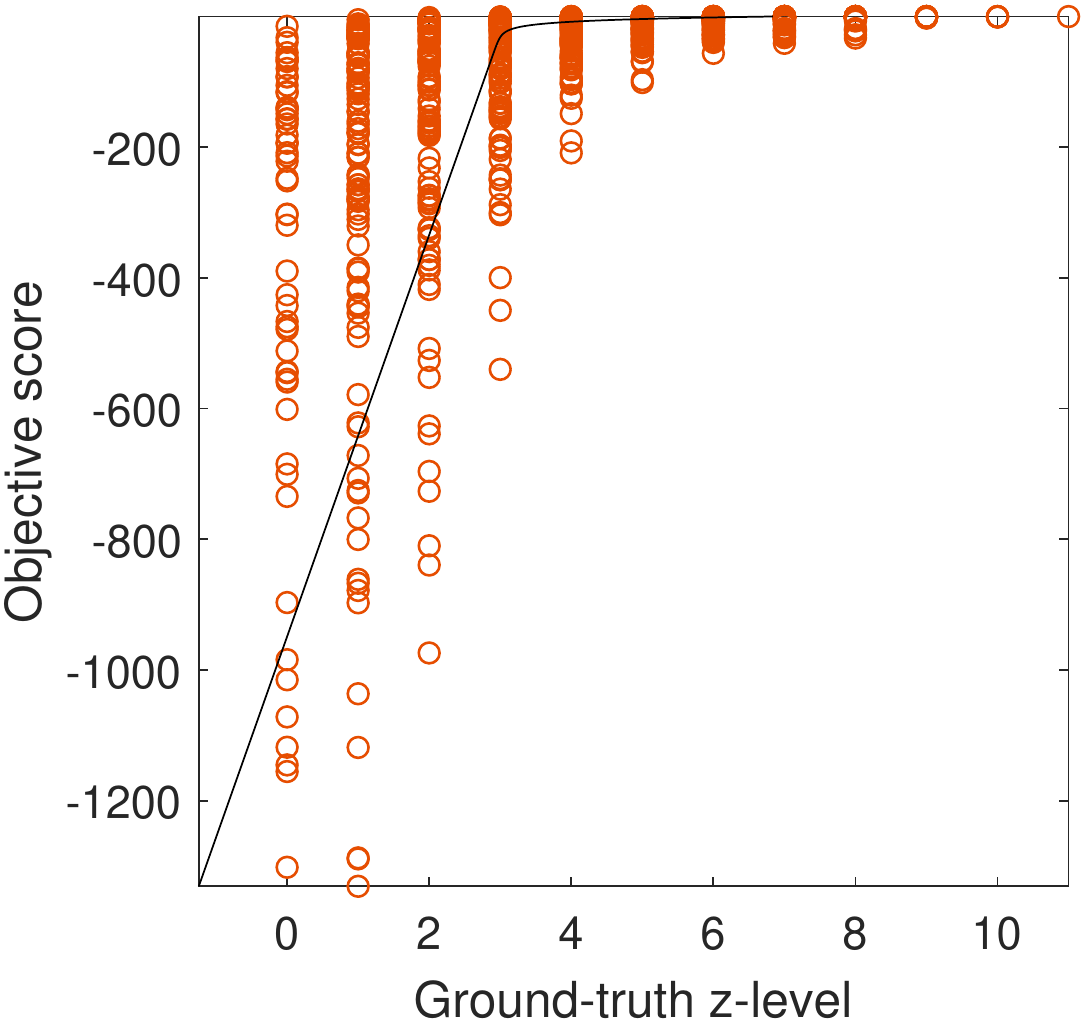}}
}\vspace{-.05in}
\centerline{
\subfigure[SPARISH \cite{li2016image}]{\includegraphics[height=0.12\textwidth]{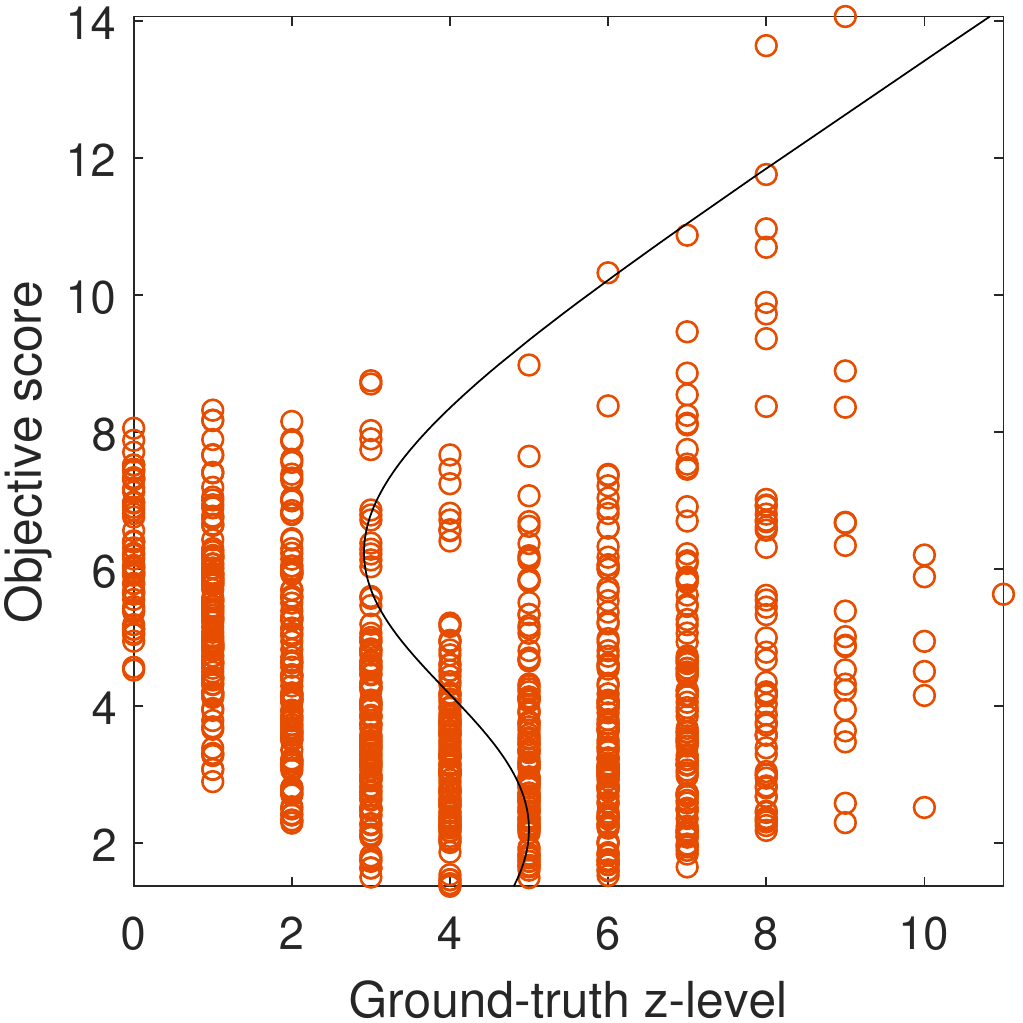}}
\subfigure[RISE \cite{li2017no}]{\includegraphics[height=0.12\textwidth]{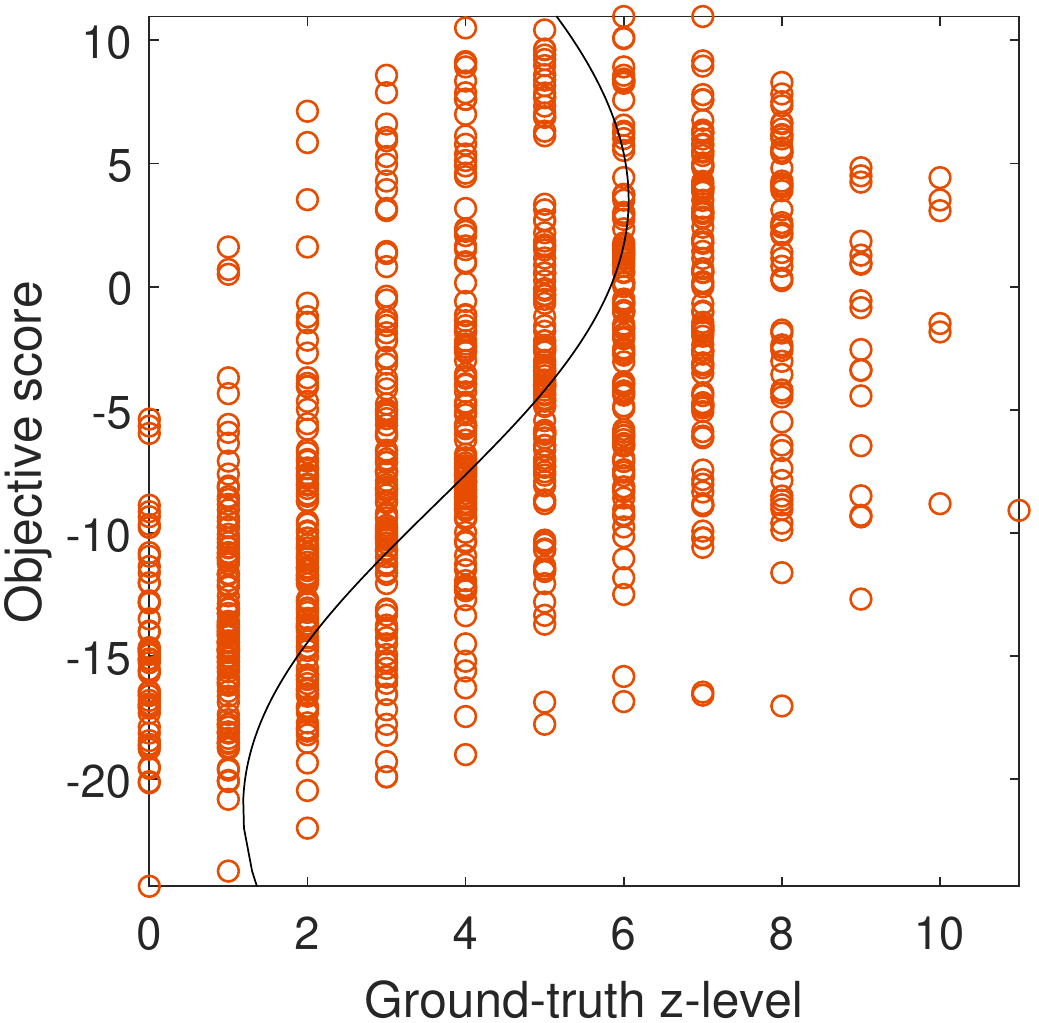}}
\subfigure[MaxPol \cite{mahdi2018image}]{\includegraphics[height=0.12\textwidth]{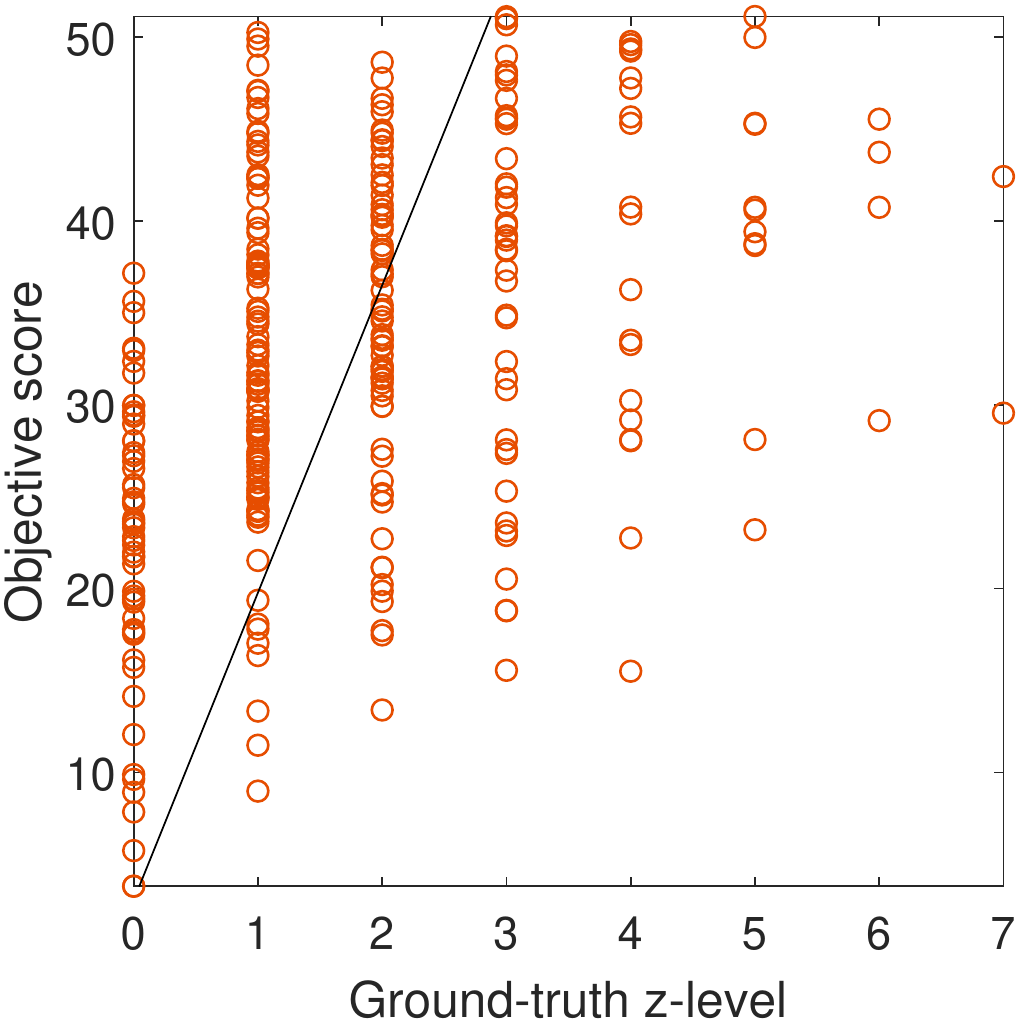}}
\subfigure[HVS-1 \cite{hosseini2018focus}]{\includegraphics[height=0.12\textwidth]{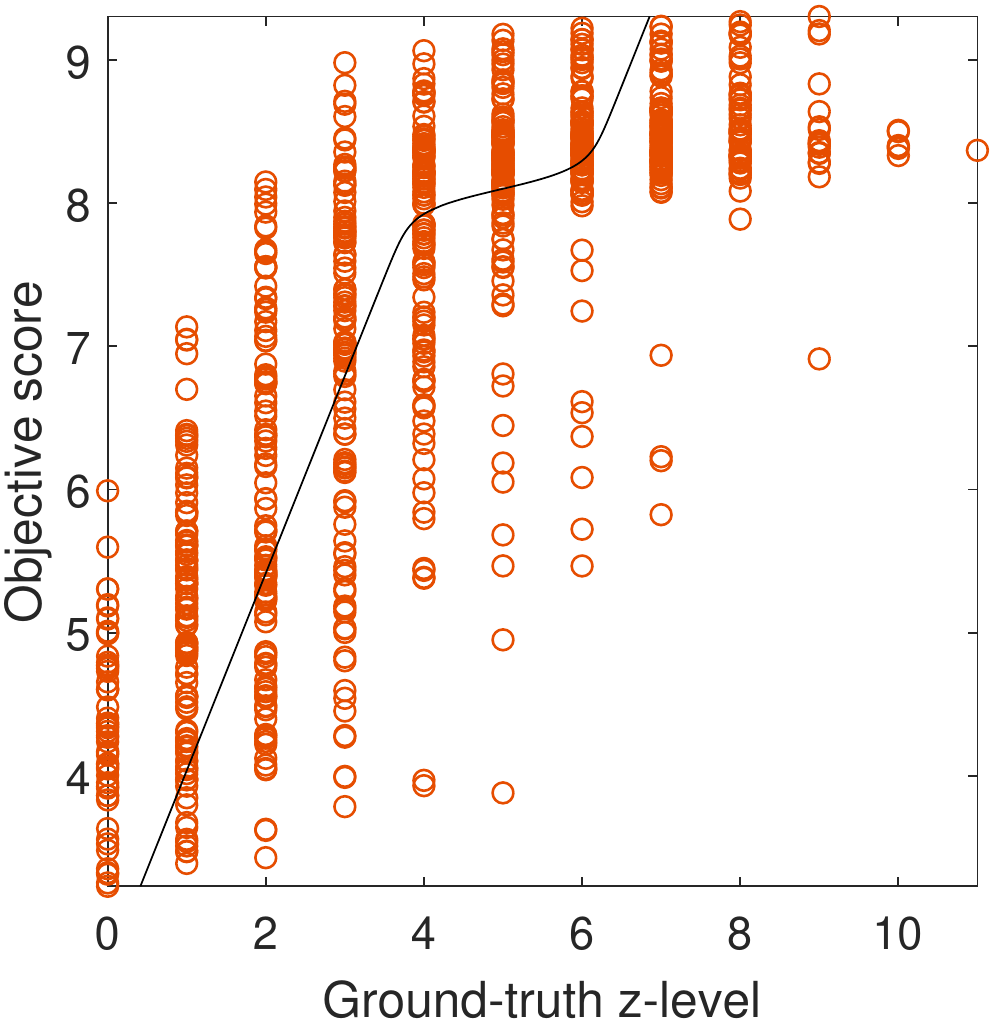}}
}\vspace{-.05in}
\centerline{
\subfigure[HVS-2 \cite{hosseini2018focus}]{\includegraphics[height=0.12\textwidth]{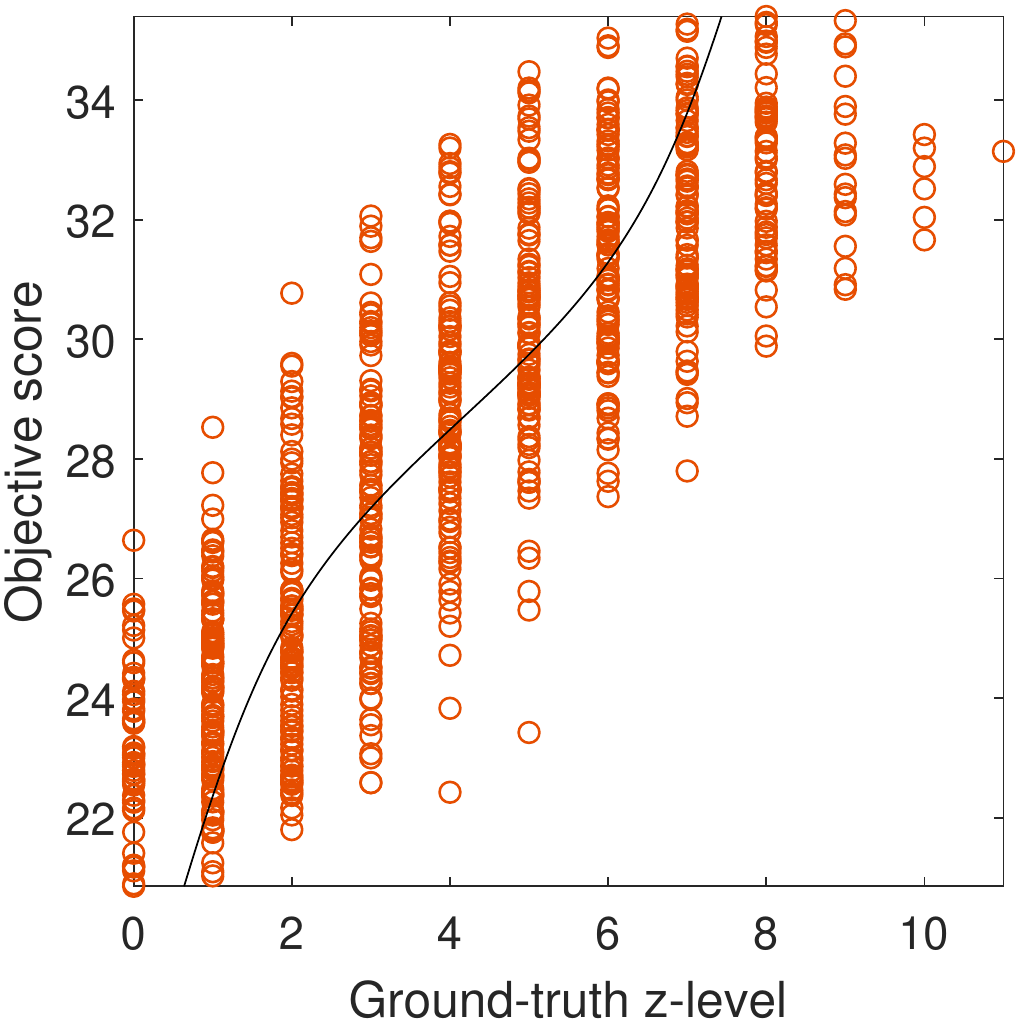}}
\subfigure[MIFQ \cite{yang2018assessing}]{\includegraphics[height=0.12\textwidth]{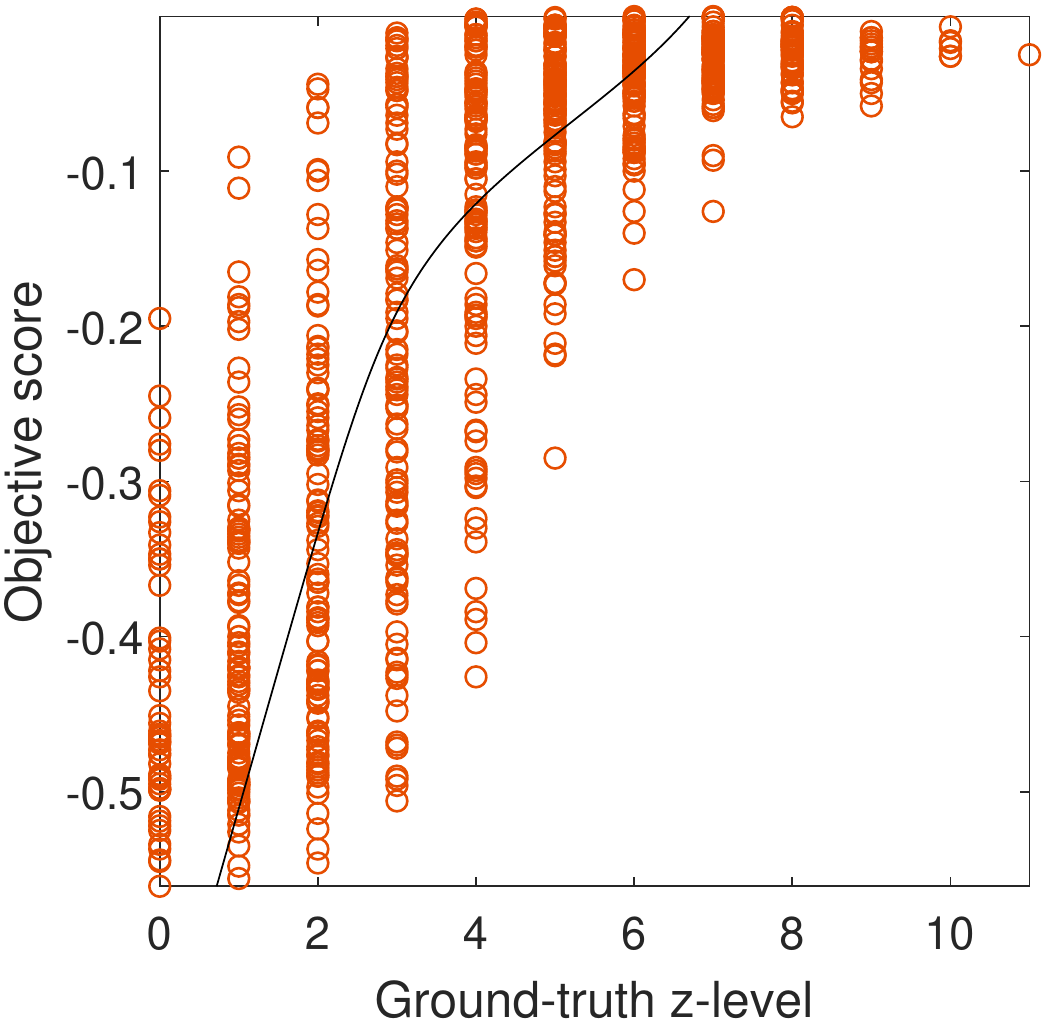}}
\subfigure[FQPath]{\includegraphics[height=0.12\textwidth]{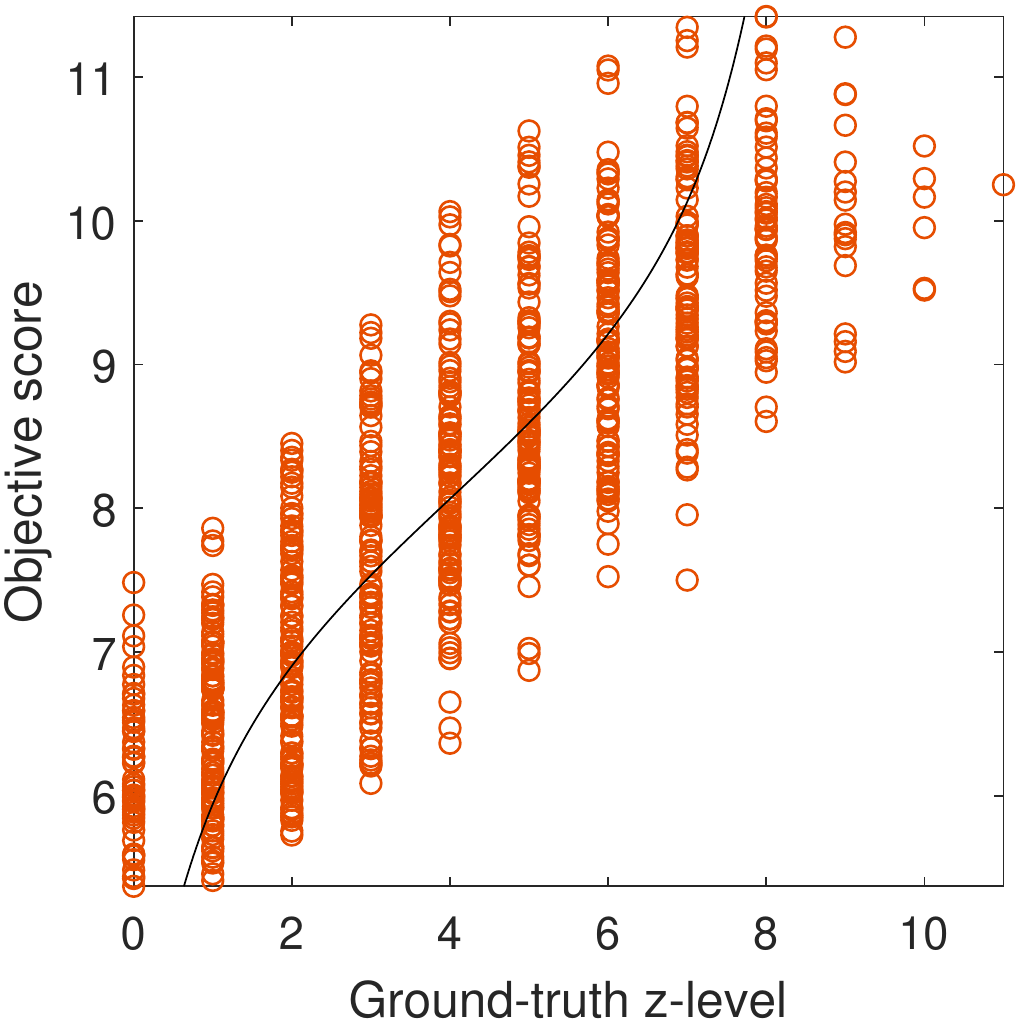}}
}
\centering
\caption{Plots of each competing focus score with the absolute value of their associated ground-truth $z$-level for each image patch in FocusPath. Note how the plot for FQPath shows a consistent linear trend between $z$-level and score which is not paralleled by the other metrics.}
\label{score_scatter}
\end{figure}

We perform the accuracy analysis on the regular version of the digital pathology database FocusPath\footnote{\url{https://sites.google.com/view/focuspathuoft}} introduced in \cite{hosseini2018focus}. The database contains 864 naturally-blurred patches (sized 1024$\times$1024 px) extracted from whole-slide scans. All 11 candidate metrics are assessed to determine their correlation with the known ground-truth $z$-level of each patch, which is measured using the Pearson linear correlation coefficient (PLCC), Spearman rank order correlation (SRCC), Kendall rank correlation coefficient (KRCC) and root mean square error (RMSE). For more information on these four accuracy measures, please refer to \cite{wang2004image, sheikh2006statistical}. PLCC measures linear correlation, SRCC measures the strength and direction of monotonicity, KRCC indicates the ordinal association, and RMSE measures the difference in value. MIFQ by default operates on patches sized 84$\times$84 (a variant of deep-learning solution), so for a fair comparison, each FocusPath patch was divided into 84$\times$84 smaller patches and their average was computed as the overall patch-level focus score. The results of the accuracy measures are provided in Table \ref{table_performance}. The top-three methods for each accuracy measure are shown in bold. FQPath achieves the highest overall performance across all four different measurements, ahead of the next best metrics HVS-2 and MIFQ. When the scores are plotted relative to the absolute value of the ground-truth $z$-levels, as in Figure \ref{score_scatter}, it is clear that FQPath is by far the most correlated with the ground-truth.

\begin{table}[htp]
	\label{table_performance}
	\caption{plcc, srcc, krcc, and rmse performance of different FQA metrics on FocusPath database. The top three metrics for each accuracy measure are shown in bold. The statistical significance (SS) of PLCC, SRCC, and KRCC accuracy results are also shown on right column. '+1' indicates that FQPath is significantly more accurate, '-1' indicates FQPath is significantly less accurate, and '0' indicates that there is no significant difference.}
	\centering
	\scriptsize
	\begin{tabular}{l|c|c|c|c|c}
	\hlinewd{1.5pt}
	Methods & PLCC & SRCC & KRCC & RMSE & SS  \\ \hline
	$\text{S}_3$ \cite{vu2012bf} & 0.7906 & 0.7914 & 0.6181 & 1.5141 & 0 \\ \hline
	MLV \cite{bahrami2014fast} & 0.3201 & 0.3296 & 0.2347 & 2.3409 & +1 \\ \hline
	$\text{ARISM}_{\text{C}}$ \cite{gu2015no} & 0.2263 & 0.3043 & 0.2195 & 2.4068 & +1 \\ \hline
	GPC \cite{leclaire2015no} & 0.7499 & 0.7811 & 0.6020 & 1.6346 & 0 \\ \hline
	SPARISH \cite{li2016image} & 0.3459 & 0.3566 & 0.2626 & 2.3184 & +1 \\ \hline
	RISE \cite{li2017no} & 0.6509 & 0.6566 & 0.4903 & 1.8758 & +1 \\ \hline
	MaxPol \cite{mahdi2018image} & 0.7056 & 0.7191 & 0.5910 & 1.7508 & 0 \\ \hline
	HVS-MaxPol-1 \cite{hosseini2018focus} & 0.8212 & 0.8144 & 0.6383 & 1.4100 & 0 \\ \hline
	HVS-MAxPol-2 \cite{hosseini2018focus} & \textbf{0.8538} & \textbf{0.8574} & \textbf{0.6852} & \textbf{1.2865} & 0 \\ \hline
	MIFQ \cite{yang2018assessing} & \textbf{0.8286} & \textbf{0.8200} & \textbf{0.6397} & \textbf{1.3834} & 0 \\ \hline
	FQPath (proposed) & \textbf{0.8556} & \textbf{0.8606} & \textbf{0.6888} & \textbf{1.2789} & N/A\\ \hlinewd{1.5pt}
	\end{tabular}
\end{table}

\subsection{Significance of Accuracy Correlation}
To determine the statistical significance of the PLCC and SRCC accuracy results, we also conducted a one-sided T-test with a $95\%$ confidence level between the proposed metric and all other metrics. As shown in Table \ref{table_performance} (right column), the proposed metric is significantly more accurate than 4 other metrics and not significantly different from the other 6 metrics. The overall results from both the accuracy and significance tests indicate that the proposed FQPath metric outranks the competing methods and provides a reliable FQ measure for WSI quality control at the patch level.

\subsection{Computational Complexity Analysis} \label{sec:experiment_complexity}
Another characteristic of a good focus quality metric for digital pathology is a low computational complexity, since digital pathology images are usually extremely large in size. For example, a 1cm$\times$1cm tissue specimen scanned at 40X resolution is 4.8 GB in size. We designed and executed two experiments for computational complexity analysis: (1) to compare the times required by the competing metrics for images of different sizes, and (2) to compare the competing metrics by their PLCC accuracy and their run-time per pixel (on 2048$\times$2048 images). Both experiments are done on a Windows station with an AMD FX-8370E 8-Core CPU 3.30 GHz.

\begin{figure}[htp]
	\centerline{
	\includegraphics[width=0.25\textwidth]{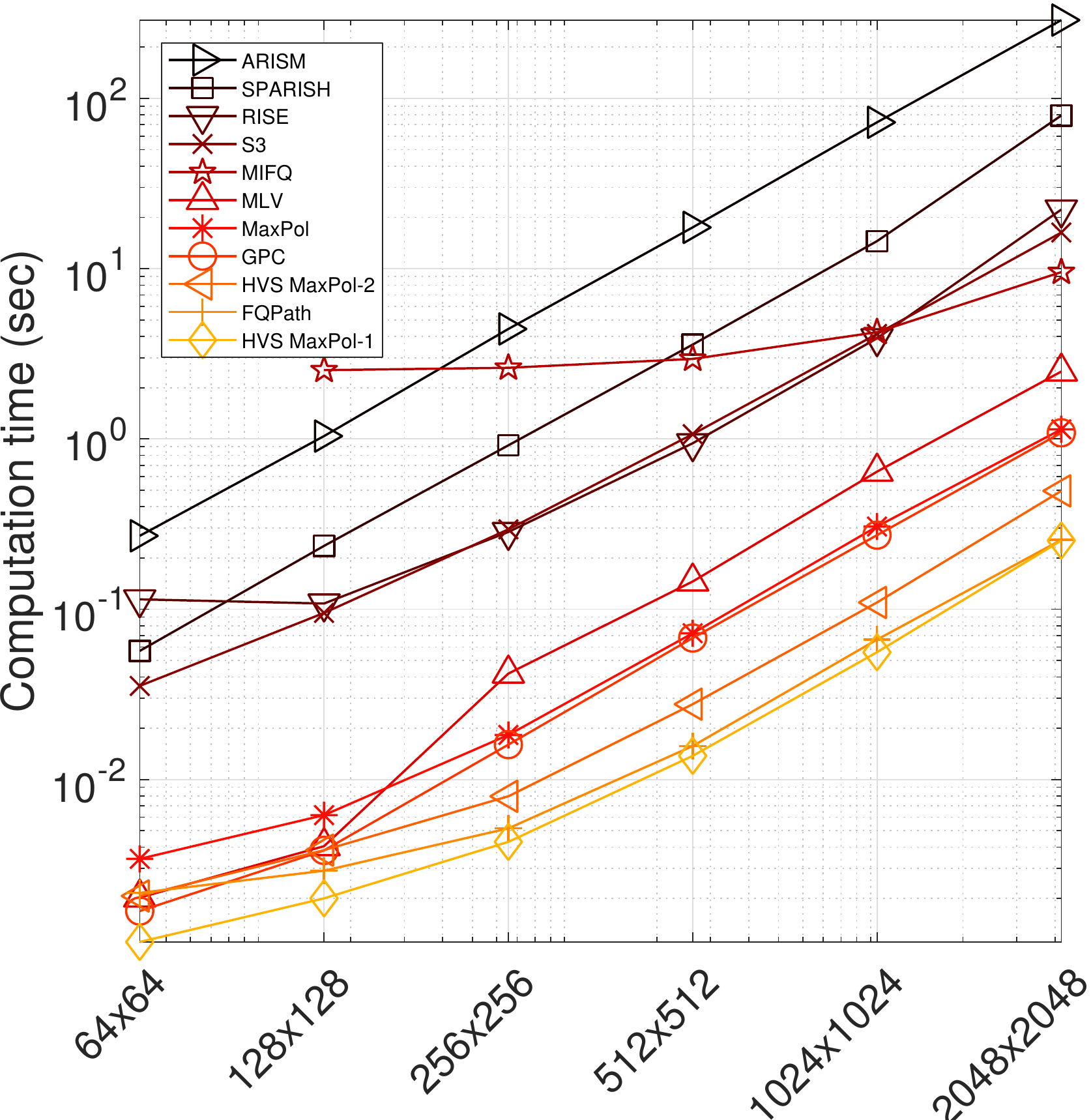}
	\includegraphics[height=0.25\textwidth]{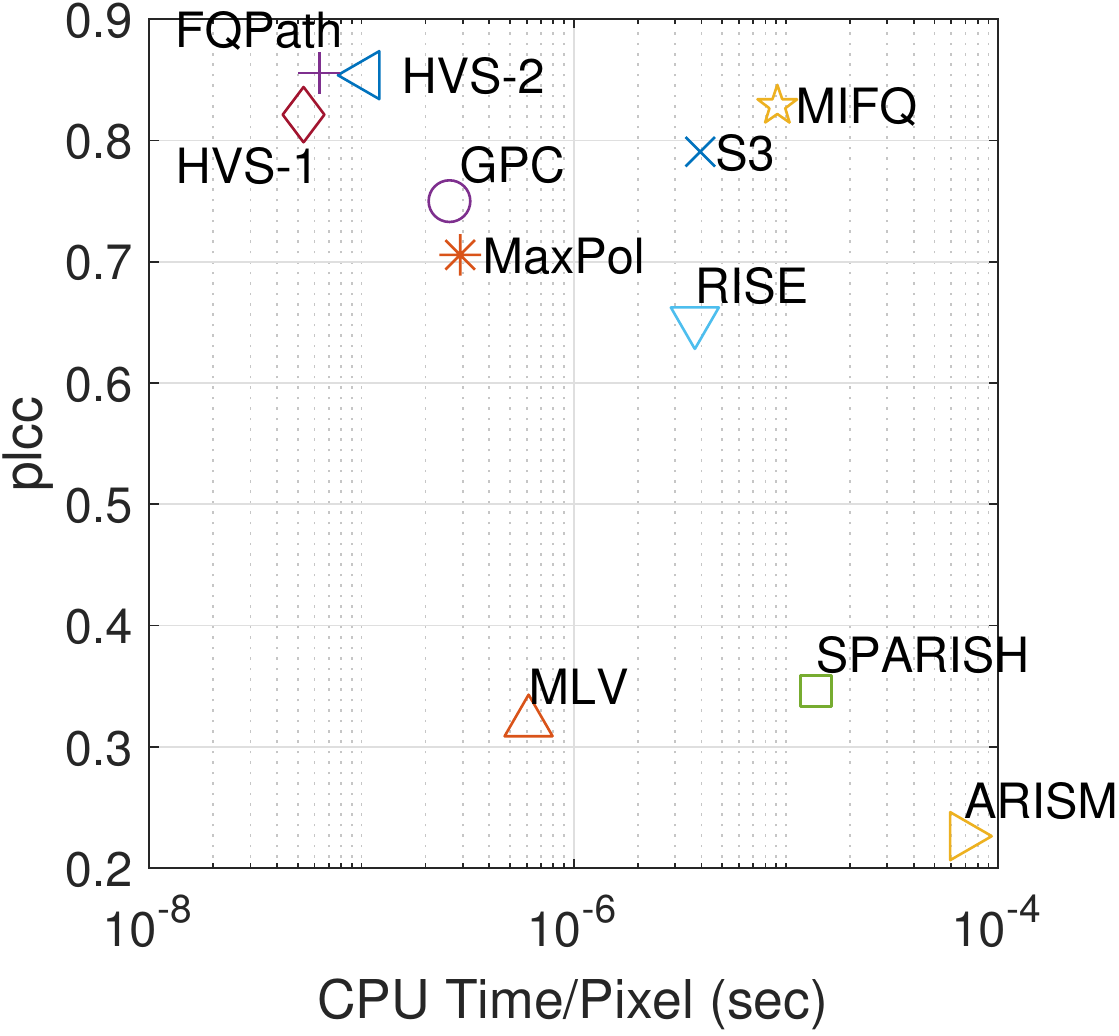}
	}\vspace{-.05in}
	\caption{Left plot: Computational complexity analysis using run time vs. image size for different FQ methods. Right plot: PLCC vs. run time per pixel using different methods. Ideal methods (high accuracy, high computational speed) are found at the top-left corner.}
	\label{fig_computational_complexity}
\end{figure}

Figure \ref{fig_computational_complexity} (left plot) demonstrates the computational complexity of all 11 FQ metrics with respect to different image sizes. We selected 20 sample images of different patch sizes i.e. $\{64, 128, 256, 512, 1024, 2048\}$ and average the the CPU time over all 20 trials to report the computation time. Note that the test rank of HVS-MaxPol-1 and FQPath are consistent over different image sizes. For a $1024\times 1024$ patch, FQPath takes $0.06626$ s, HVS-MaxPol-2 takes double the time ($0.1905$ s), and MIFQ takes 74 times ($4.234$ s).


Figure \ref{fig_computational_complexity} (right plot) demonstrates the relationship between PLCC performance and CPU run time. A large y- (vertical) axis value indicates high accuracy performance and low x- (horizontal) axis value indicates high speed. Therefore, an ideal method would be located at the top-left corner, such as FQPath, HVS-MaxPol-1, and HVS-MaxPol-2. Overall, FQPath performs excellently in terms of both accuracy and speed, proving to be highly reliable for digital pathology FQ assessment.
\section{Experiment No.2: WSI-Level Analysis}
\label{sec:experiment_heatmap}
At the whole-slide level, FQPath can be applied (with the inverse Gaussian projection) to construct a slide focus quality heatmap and visualize the regions of high and low focus quality. In this section, we propose two experiments to show the validity of our approach to whole-slide-level focus quality assessment: (1) display the FQ heatmaps generated for some select whole slide images, and (2) correlate the FQ heatmaps with the subjective slide quality scores.

\begin{table}[htp]
\renewcommand{\arraystretch}{1.3}
\caption{Raw pyramid images and heat maps. (a)-(f) show the raw pyramid images, (g)-(l) show the corresponding focus quality heatmaps, and (m)-(r) show the corresponding heatmap score cumulative summation curves (faster-rising curves indicating better overall slide quality). A 10-bin histogram of all the heatmaps is taken and one example patch taken from each bin of the slide (indicated by the red arrows, with larger bin numbers indicating better focus quality). See the patches themselves inside Figure \ref{slide_image_patch}.}
\label{heatmaps}\vspace{-.1in}
\begin{center}
\begin{tabular}{|c|c|c|c|c|c|}
\cline{2-6}
\multicolumn{1}{c|}{} & Slide-1 & Slide-2 & Slide-3 & Slide-4  & Slide-5 \\
\cline{1-6} 
{\hspace{-.05in}\begin{sideways} \hspace{.4in}WSI \end{sideways}\hspace{-.05in}} &
{\hspace{-.05in}\includegraphics[width=0.075\textwidth]{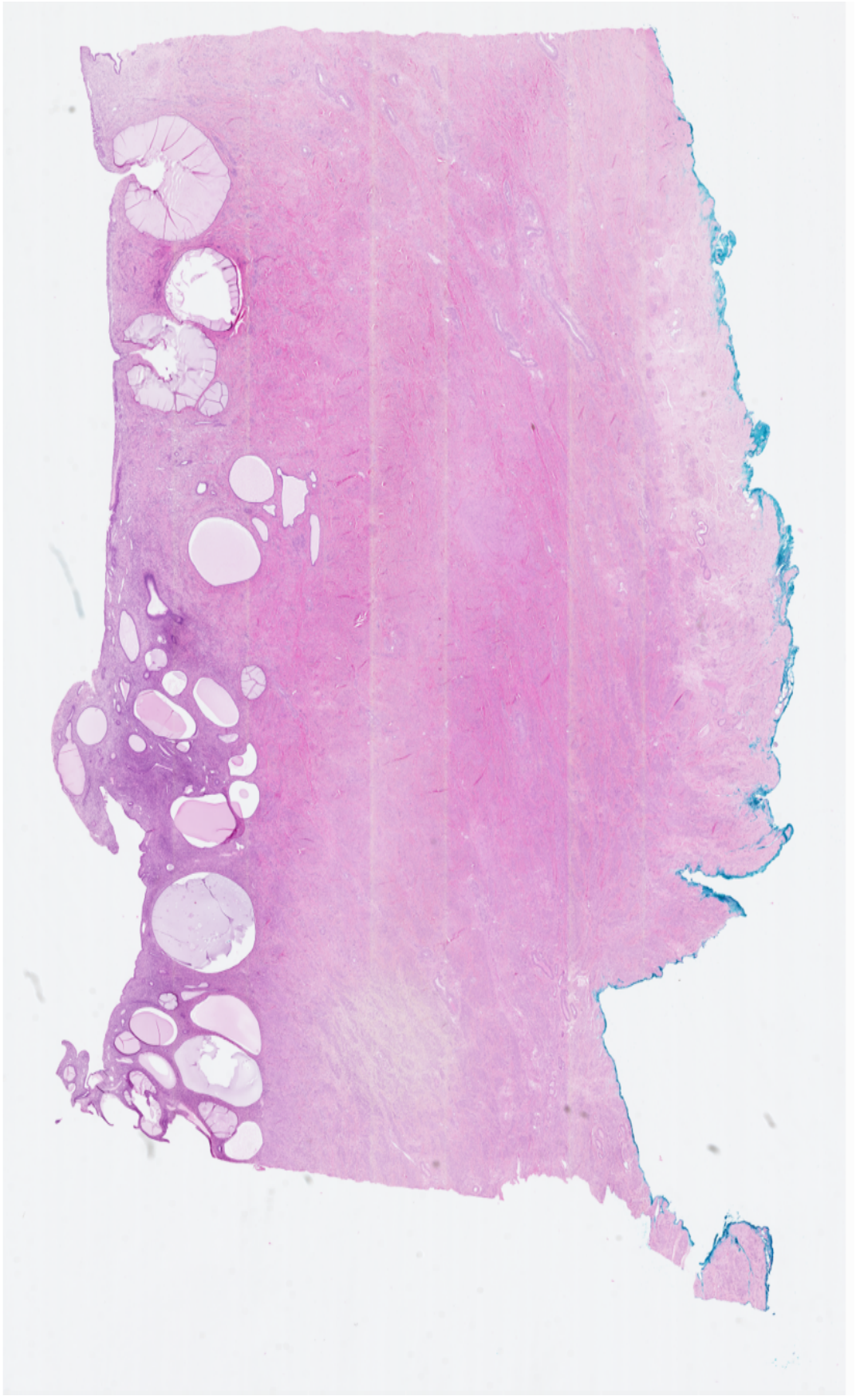}\hspace{-.05in}} &
{\hspace{-.05in}\includegraphics[width=0.075\textwidth]{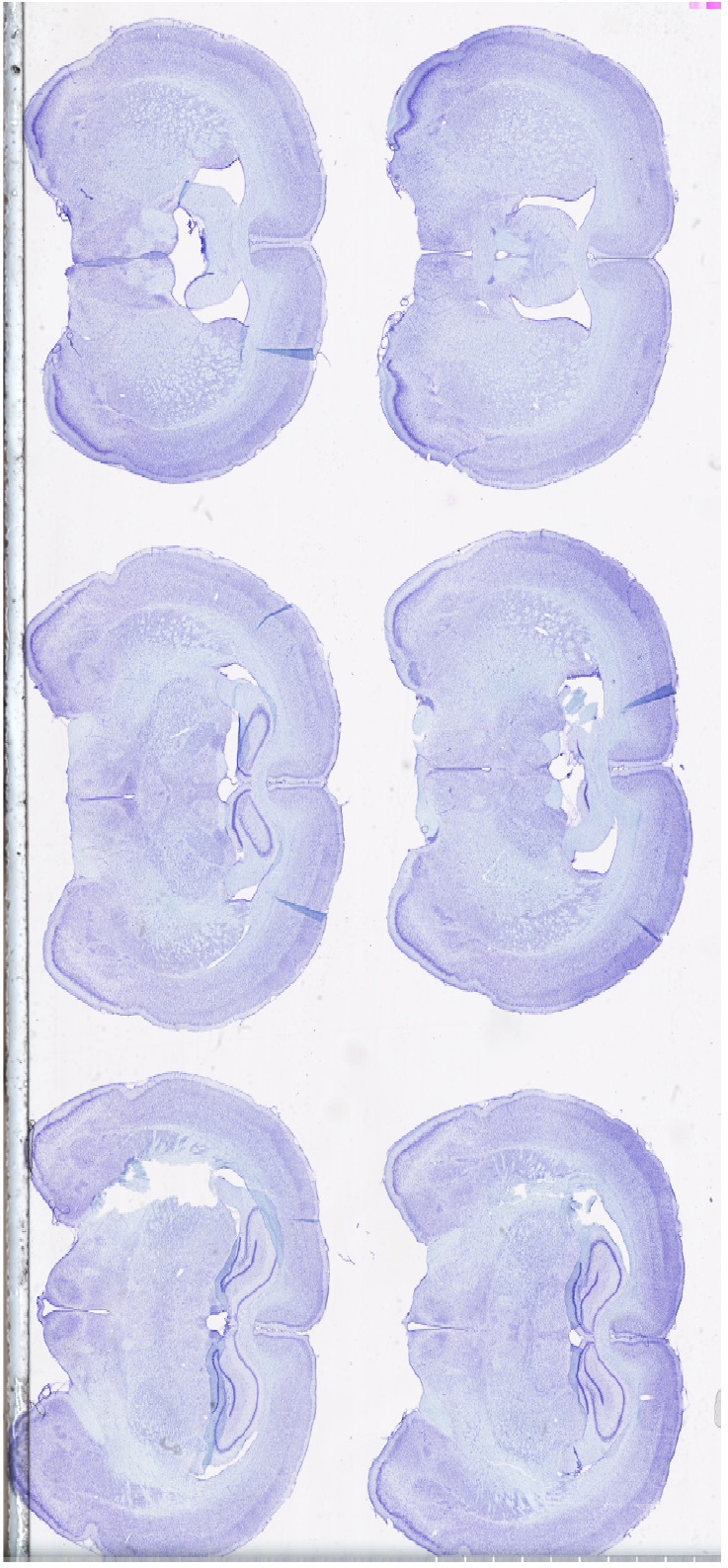}\hspace{-.05in}} &
{\hspace{-.05in}\includegraphics[width=0.075\textwidth]{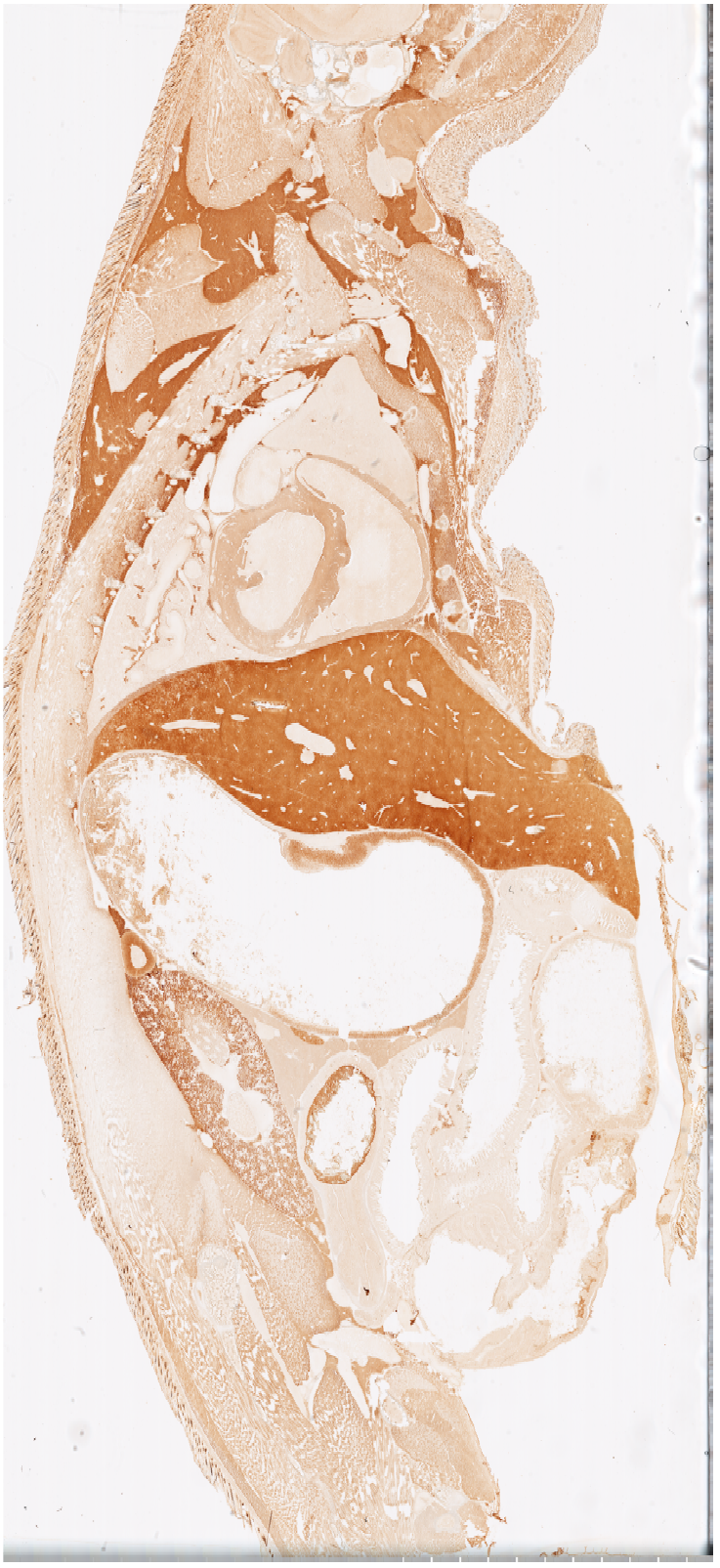}\hspace{-.05in}} &
{\hspace{-.05in}\includegraphics[width=0.075\textwidth]{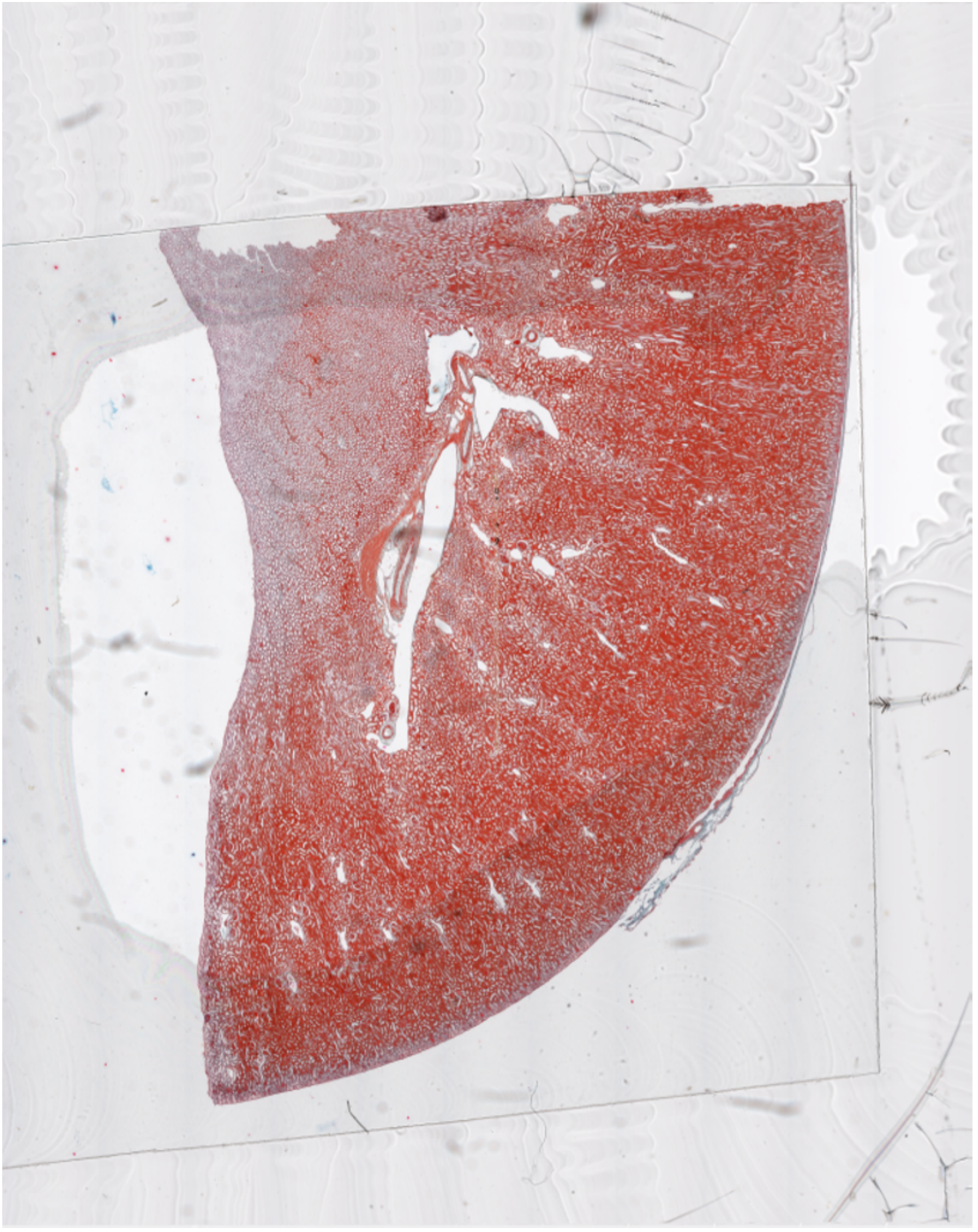}\hspace{-.05in}} &
{\hspace{-.05in}\includegraphics[width=0.075\textwidth]{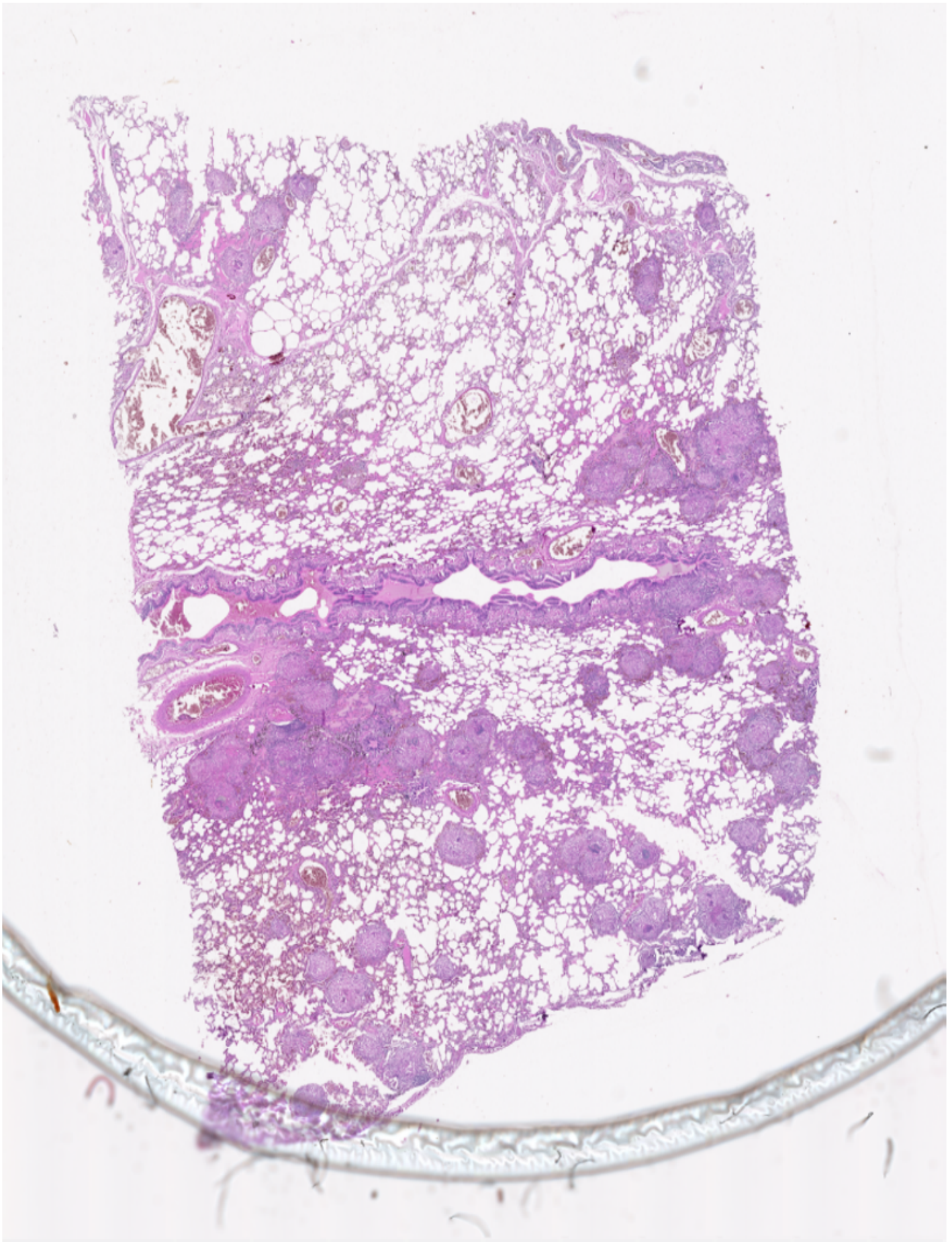}\hspace{-.05in}} \\
\hline
{\hspace{-.05in}\begin{sideways} \hspace{.3in}FQ-Heatmap \end{sideways}\hspace{-.05in}} &
{\hspace{-.05in}\includegraphics[width=0.075\textwidth]{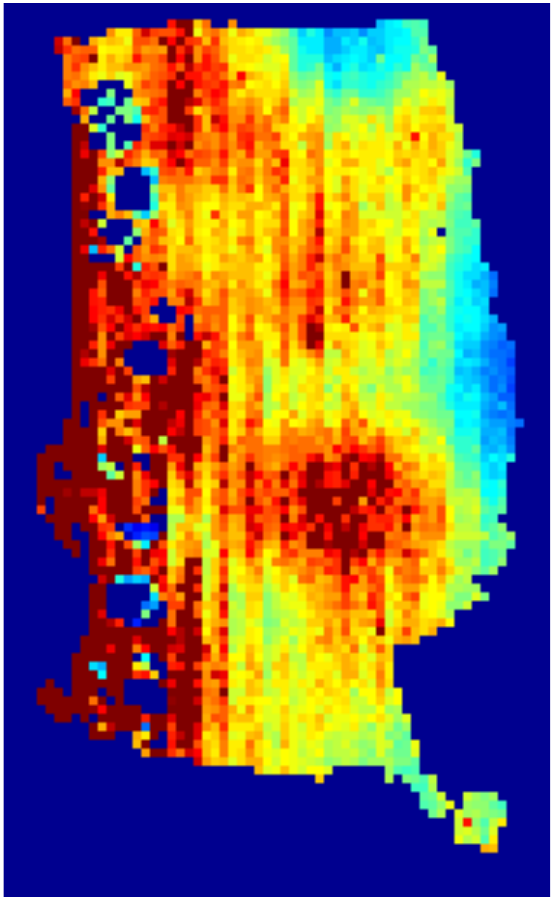}\hspace{-.05in}} &
{\hspace{-.05in}\includegraphics[width=0.075\textwidth]{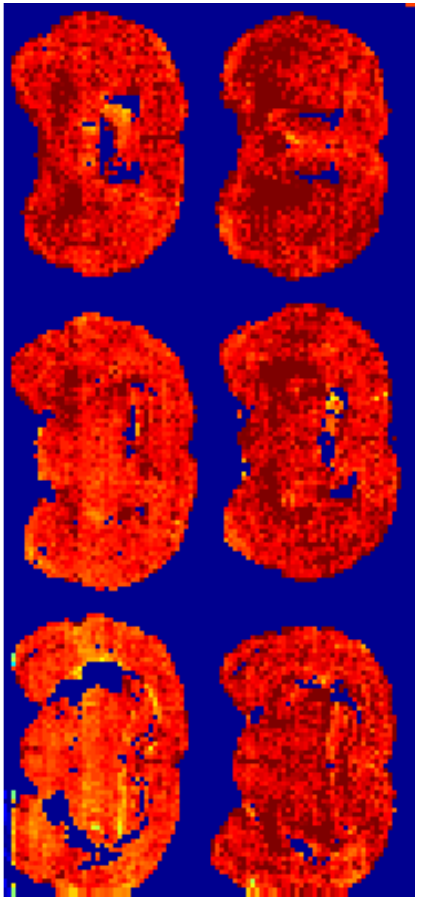}\hspace{-.05in}} &
{\hspace{-.05in}\includegraphics[width=0.075\textwidth]{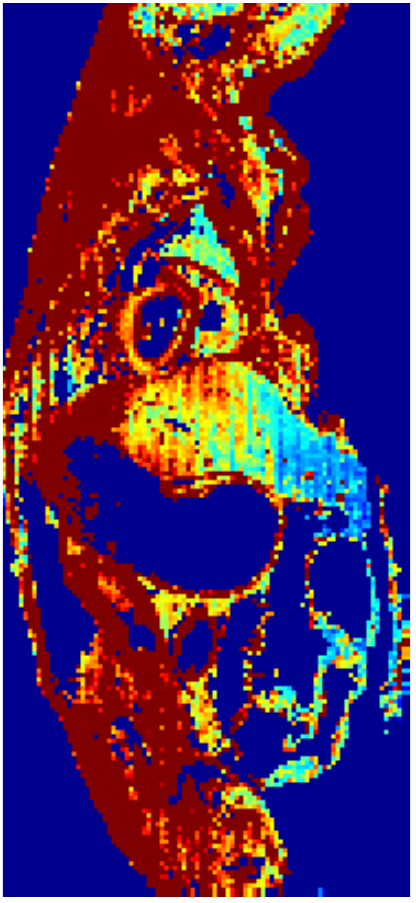}\hspace{-.05in}} &
{\hspace{-.05in}\includegraphics[width=0.075\textwidth]{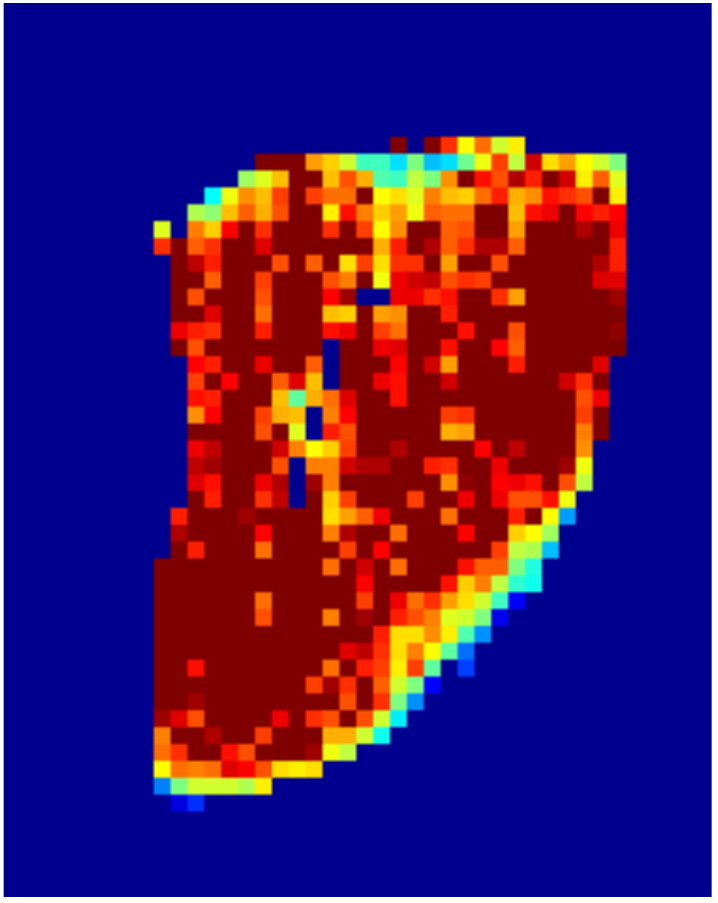}\hspace{-.05in}} &
{\hspace{-.05in}\includegraphics[width=0.075\textwidth]{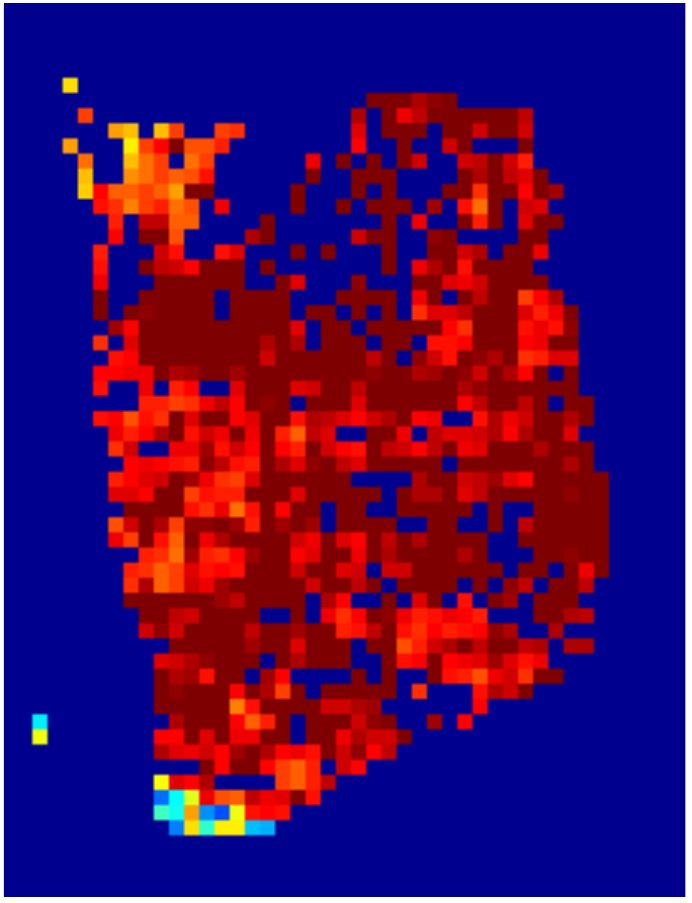}\hspace{-.05in}} \\
\hline
{\hspace{-.05in}\begin{sideways} \hspace{.0in}Cumsum Curve \end{sideways}\hspace{-.05in}} &
{\hspace{-.05in}\includegraphics[width=0.075\textwidth]{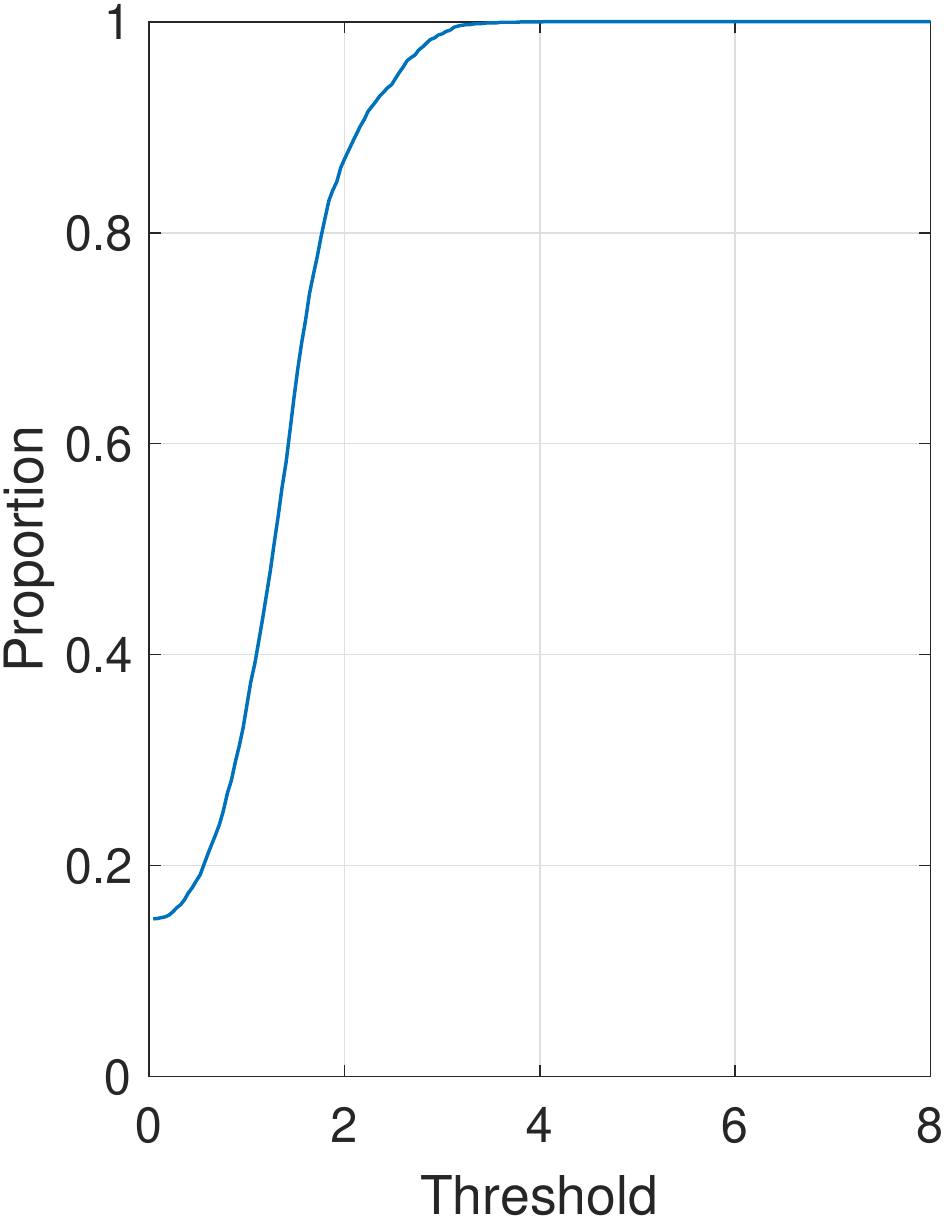}\hspace{-.05in}} &
{\hspace{-.05in}\includegraphics[width=0.075\textwidth]{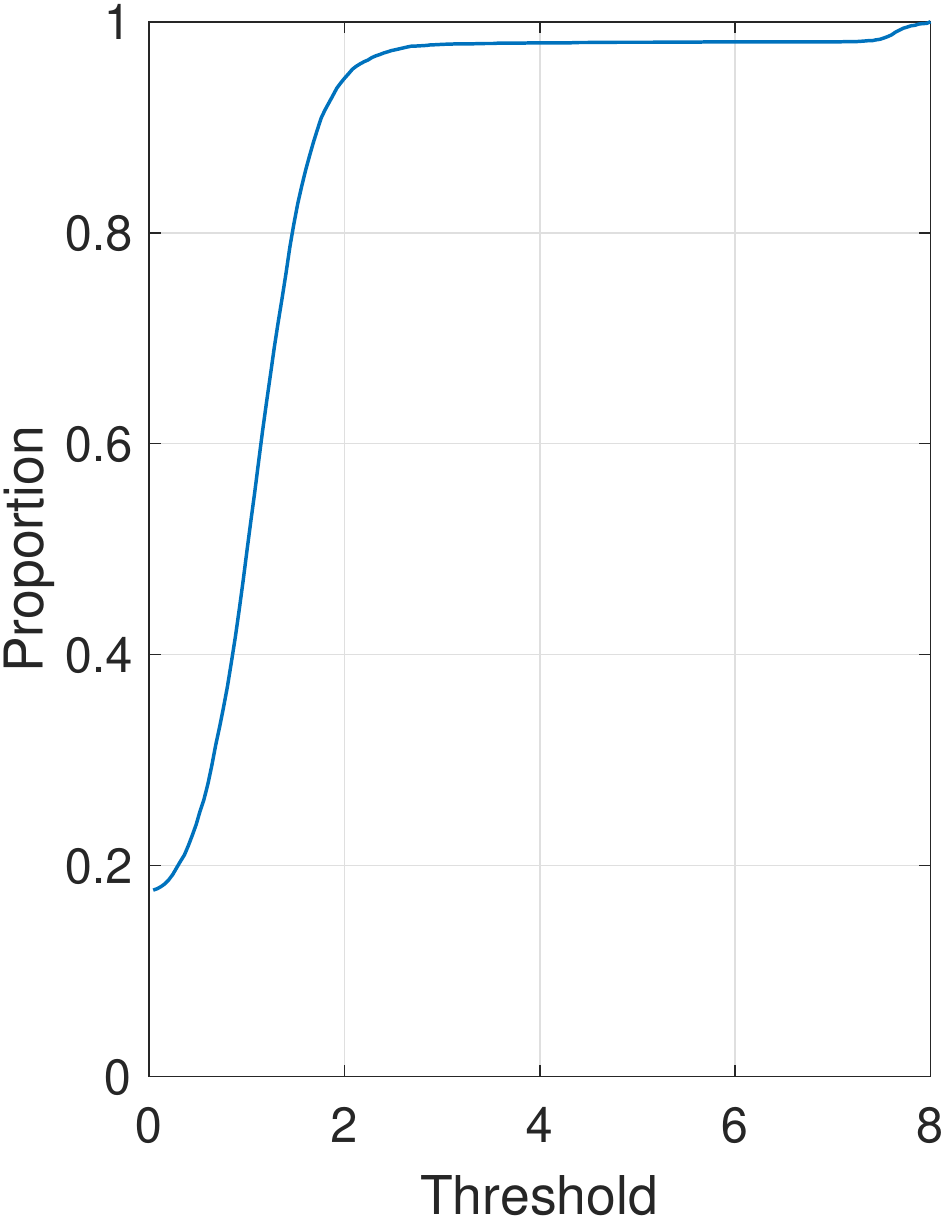}\hspace{-.05in}} &
{\hspace{-.05in}\includegraphics[width=0.075\textwidth]{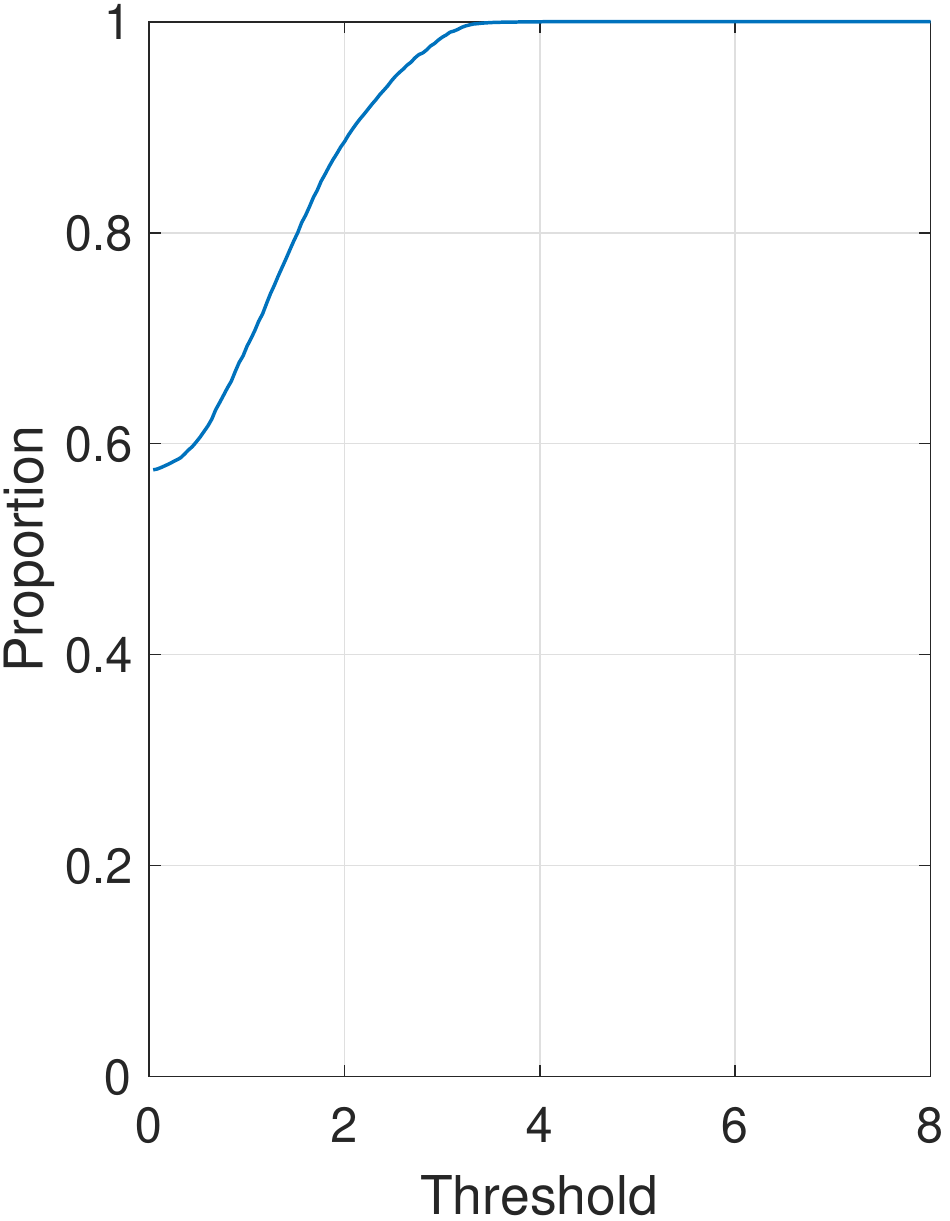}\hspace{-.05in}} &
{\hspace{-.05in}\includegraphics[width=0.075\textwidth]{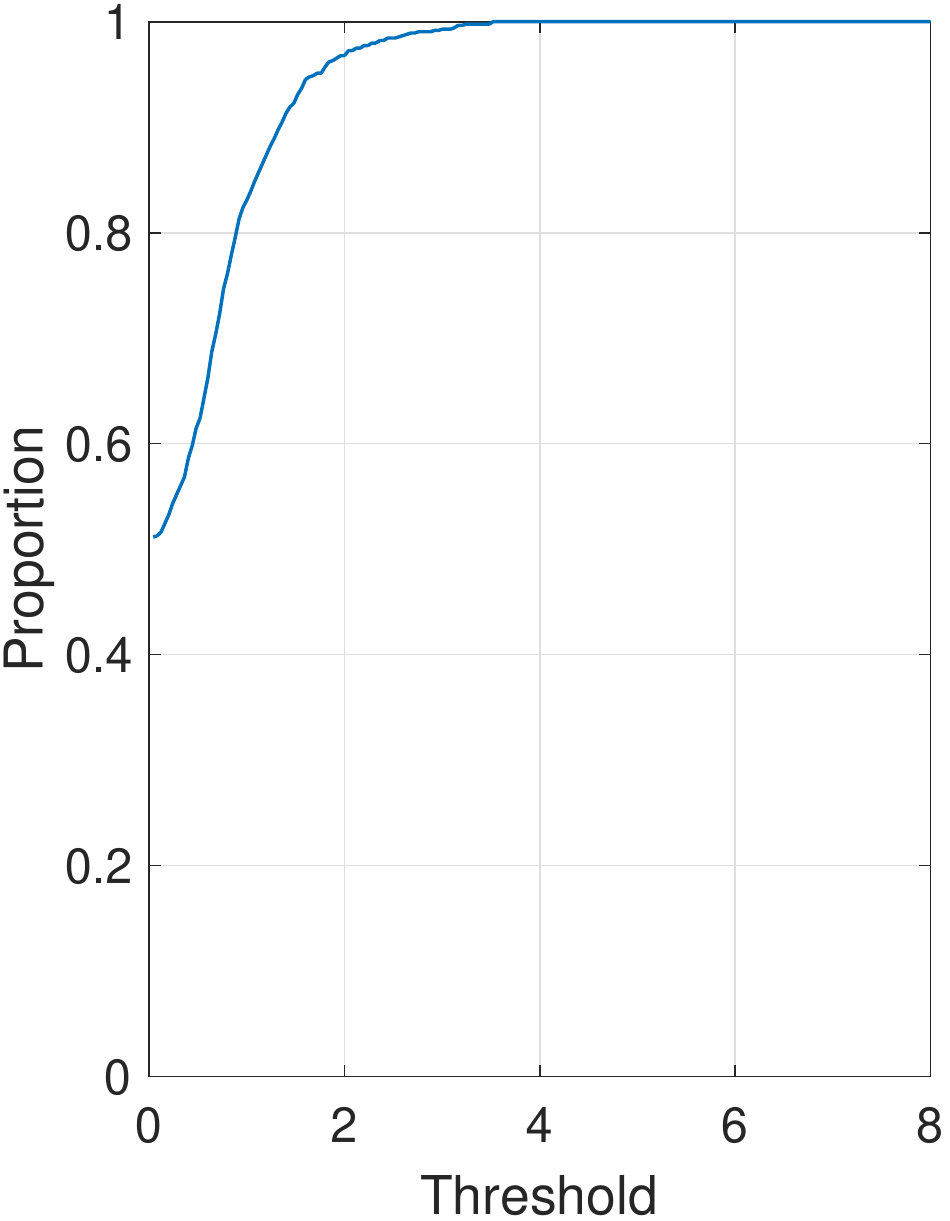}\hspace{-.05in}} &
{\hspace{-.05in}\includegraphics[width=0.075\textwidth]{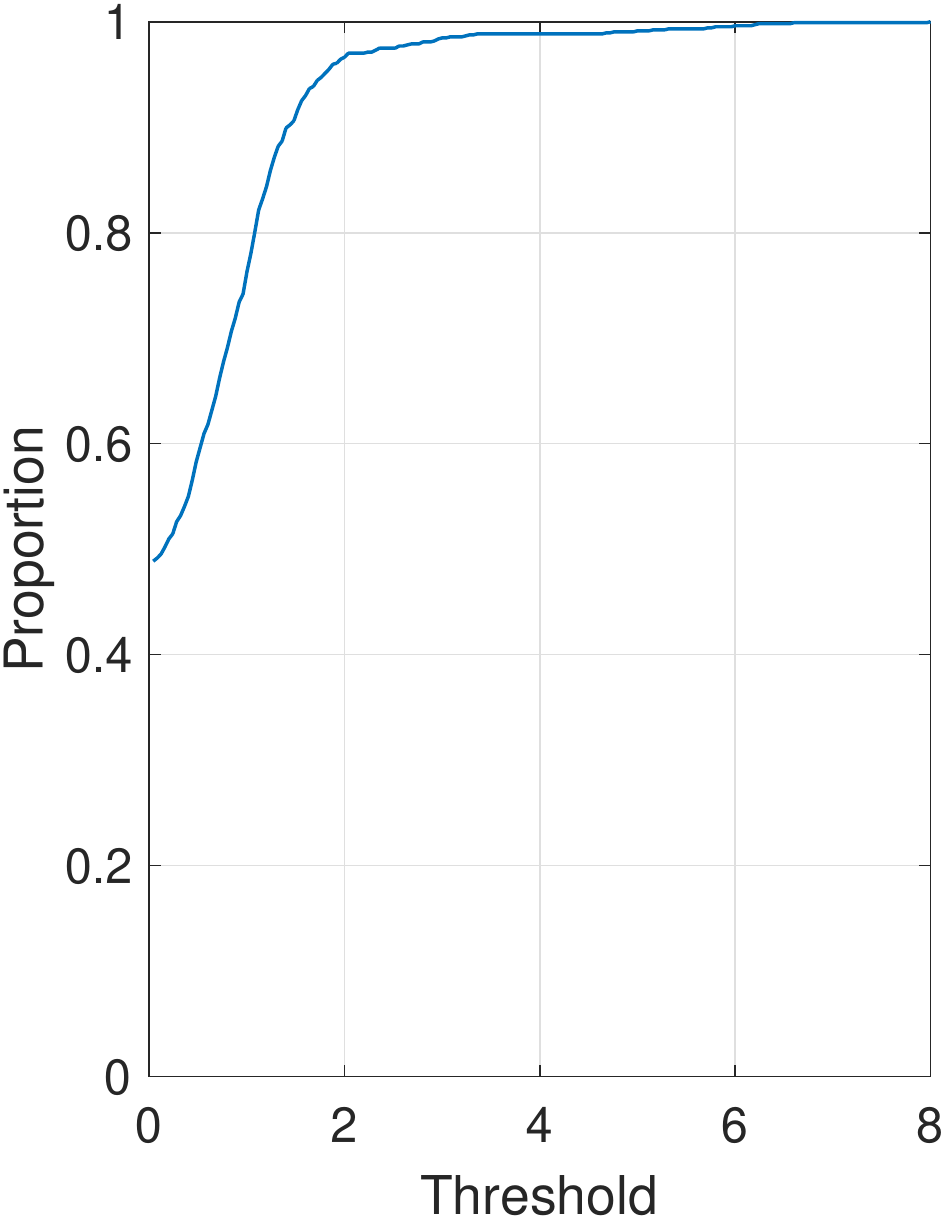}\hspace{-.05in}} \\
\hline
\end{tabular}
\end{center}
\end{table}\vspace{-.1in}

\subsection{Heat Map Analysis}
In Table \ref{heatmaps}, we show the FQ heatmaps generated for five select slides. The shade of colors in heatmaps are demonstrated from blue to red corresponding from low to high focus quality, respectively. Note how local regions of low focus quality are easily visible in shades of blue and cyan, and vertical scan artifacts in slides 1 and 3 show as stripe-like patterns. The distribution of patch quality scores in these focus quality heatmaps is indicative of overall scan quality and decide whether to accept or reject the scan.

In Table \ref{slide_image_patch}, we show example patches for each of five histogram bins for each heatmap. Note how the patches vary in focus quality from crisp tissue scans to increasing degrees of blur. This supports the validity of our proposed focus quality heatmap where the processing metric provides a reliable and robust analysis of focus measures within intraclass and interclass of tissue types across different slide images.

\begin{table}[htp]
\renewcommand{\arraystretch}{1.3}
\caption{Example image patches from five slide images.}
\label{slide_image_patch}\vspace{-.1in}
\begin{center}
\begin{tabular}{|c|c|c|c|c|c|}
\cline{2-6}
\multicolumn{1}{c|}{} & Slide-1 & Slide-2 & Slide-3 & Slide-4 & Slide-5 \\
\cline{1-6} 
{\hspace{-.05in}\begin{sideways} \hspace{.1in}Bin-1 \end{sideways}\hspace{-.05in}} &
{\hspace{-.05in}\includegraphics[width=0.075\textwidth]{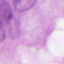}\hspace{-.05in}} &
{\hspace{-.05in}\includegraphics[width=0.075\textwidth]{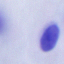}\hspace{-.05in}} &
{\hspace{-.05in}\includegraphics[width=0.075\textwidth]{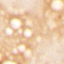}\hspace{-.05in}} &
{\hspace{-.05in}\includegraphics[width=0.075\textwidth]{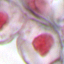}\hspace{-.05in}} &
{\hspace{-.05in}\includegraphics[width=0.075\textwidth]{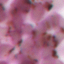}\hspace{-.05in}} \\
\hline
{\hspace{-.05in}\begin{sideways} \hspace{.1in}Bin-2 \end{sideways}\hspace{-.05in}} &
{\hspace{-.05in}\includegraphics[width=0.075\textwidth]{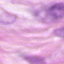}\hspace{-.05in}} &
{\hspace{-.05in}\includegraphics[width=0.075\textwidth]{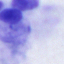}\hspace{-.05in}} &
{\hspace{-.05in}\includegraphics[width=0.075\textwidth]{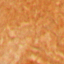}\hspace{-.05in}} &
{\hspace{-.05in}\includegraphics[width=0.075\textwidth]{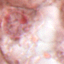}\hspace{-.05in}} &
{\hspace{-.05in}\includegraphics[width=0.075\textwidth]{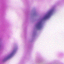}\hspace{-.05in}} \\
\hline
{\hspace{-.05in}\begin{sideways} \hspace{.1in}Bin-3 \end{sideways}\hspace{-.05in}} &
{\hspace{-.05in}\includegraphics[width=0.07\textwidth]{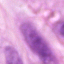}\hspace{-.05in}} &
{\hspace{-.05in}\includegraphics[width=0.07\textwidth]{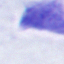}\hspace{-.05in}} &
{\hspace{-.05in}\includegraphics[width=0.07\textwidth]{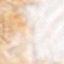}\hspace{-.05in}} &
{\hspace{-.05in}\includegraphics[width=0.07\textwidth]{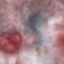}\hspace{-.05in}} &
{\hspace{-.05in}\includegraphics[width=0.07\textwidth]{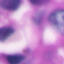}\hspace{-.05in}} \\
\hline
{\hspace{-.05in}\begin{sideways} \hspace{.1in}Bin-4 \end{sideways}\hspace{-.05in}} &
{\hspace{-.05in}\includegraphics[width=0.07\textwidth]{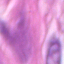}\hspace{-.05in}} &
{\hspace{-.05in}\includegraphics[width=0.07\textwidth]{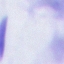}\hspace{-.05in}} &
{\hspace{-.05in}\includegraphics[width=0.07\textwidth]{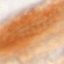}\hspace{-.05in}} &
{\hspace{-.05in}\includegraphics[width=0.07\textwidth]{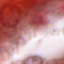}\hspace{-.05in}} &
{\hspace{-.05in}\includegraphics[width=0.07\textwidth]{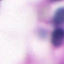}\hspace{-.05in}} \\
\hline
{\hspace{-.05in}\begin{sideways} \hspace{.1in}Bin-5 \end{sideways}\hspace{-.05in}} &
{\hspace{-.05in}\includegraphics[width=0.07\textwidth]{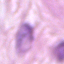}\hspace{-.05in}} &
{\hspace{-.05in}\includegraphics[width=0.07\textwidth]{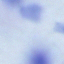}\hspace{-.05in}} &
{\hspace{-.05in}\includegraphics[width=0.07\textwidth]{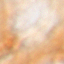}\hspace{-.05in}} &
{\hspace{-.05in}\includegraphics[width=0.07\textwidth]{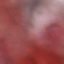}\hspace{-.05in}} &
{\hspace{-.05in}\includegraphics[width=0.07\textwidth]{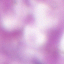}\hspace{-.05in}} \\
\hline
{\hspace{-.05in}\begin{sideways} \hspace{.1in}Bin-6 \end{sideways}\hspace{-.05in}} &
{\hspace{-.05in}\includegraphics[width=0.07\textwidth]{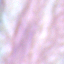}\hspace{-.05in}} &
{\hspace{-.05in}\includegraphics[width=0.07\textwidth]{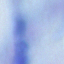}\hspace{-.05in}} &
{\hspace{-.05in}\includegraphics[width=0.07\textwidth]{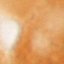}\hspace{-.05in}} &
{\hspace{-.05in}\includegraphics[width=0.07\textwidth]{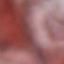}\hspace{-.05in}} &
{\hspace{-.05in}\includegraphics[width=0.07\textwidth]{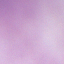}\hspace{-.05in}} \\
\hline
\end{tabular}
\end{center}
\end{table}\vspace{-.1in}

\subsection{Subjective Whole-Slide Scoring}
\begin{figure}[htp]
	\centerline{
	\includegraphics[height=0.18\textwidth]{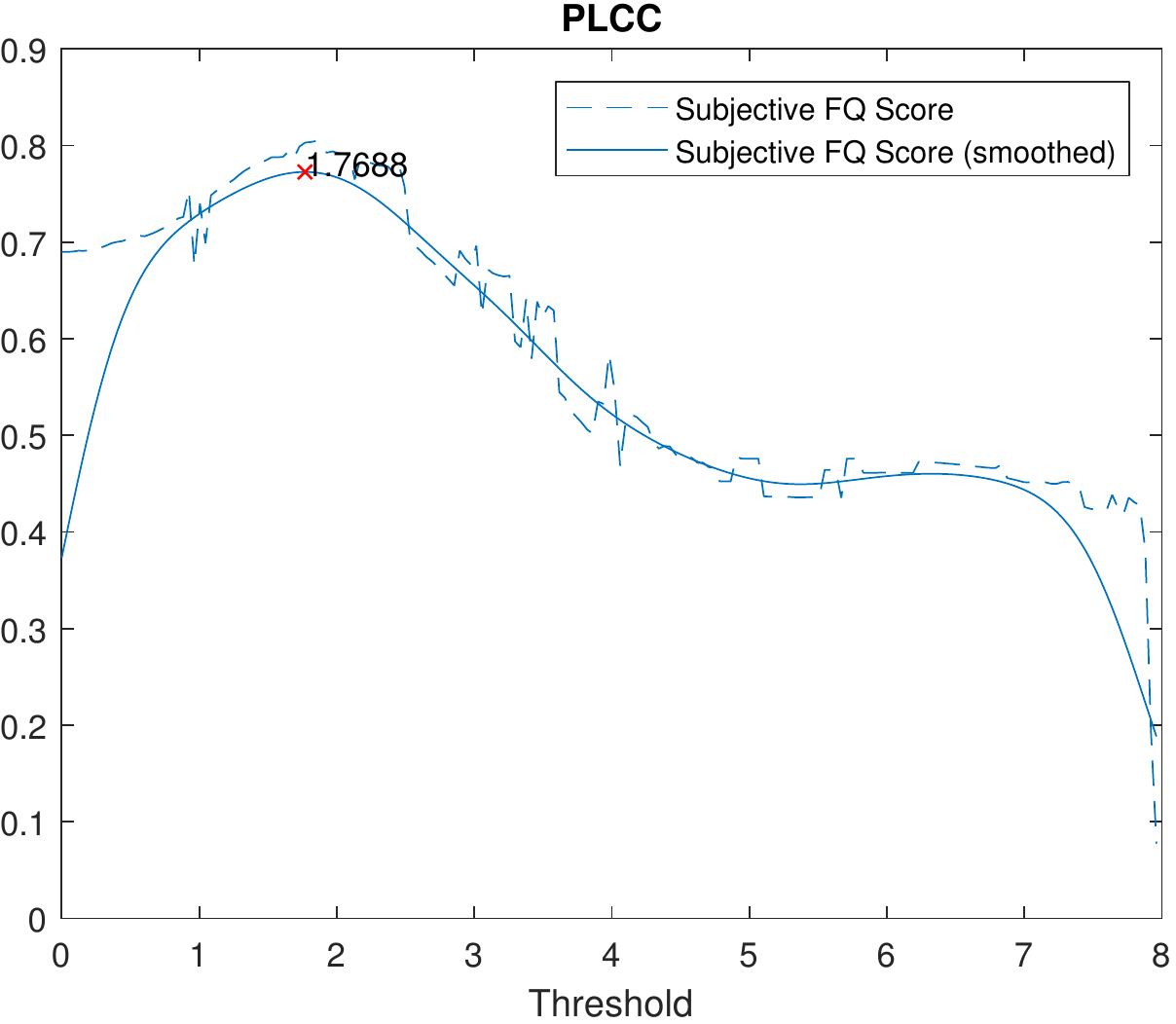}
	\includegraphics[height=0.18\textwidth]{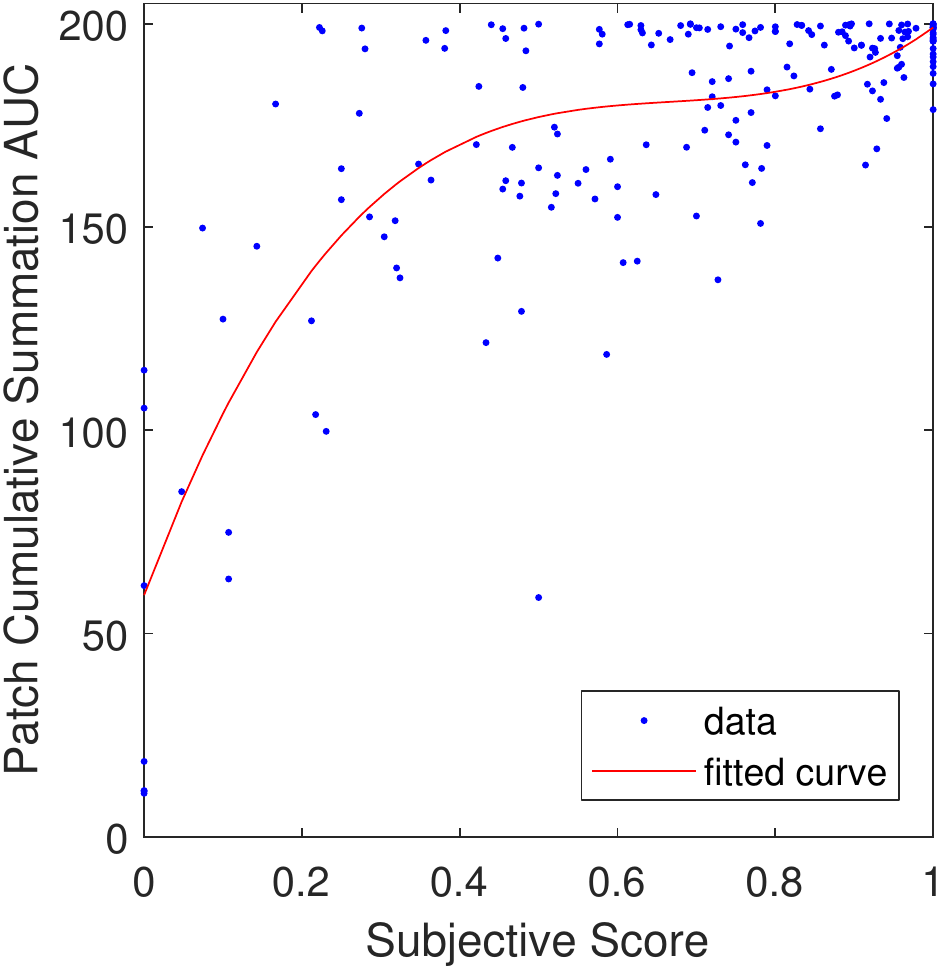}
	}
	\centering
	\caption{Left plot: PLCC across different threshold values, between the slide-level FQPath objective score and the subjective acceptance ratio. Smoothed plot shown as solid line with objective acceptance threshold of 1.7688 that maximizes correlation between the objective and subjective slide-level acceptance ratio. Right plot: Objective Acceptance Ratio (using acceptance threshold of 1.7688) vs. Subjective Acceptance Ratio. Red line indicates IQA fit through data}
	\label{Subj_PLCC}
\end{figure}

In order to correlate our FQ heatmaps with our subjective slide quality scores, we considered the heatmap patch quality scores independently and generated cumulative summation curves to determine their distribution. In Figure \ref{heatmaps} (third row), we show these cumulative summation curves for five different WSIs. In Figures \ref{Subj_PLCC} (left plot) we show the PLCC between the heatmap scores with the subjective slide scores, and determined the optimal heatmap score threshold to be 1.7688 based on the PLCC plot. And in Figure \ref{Subj_PLCC} (right plot), we plot the optimally-thresholded heatmap acceptance ratio with the subjective slide scores. As can be seen, the objective heatmap acceptance ratios form a linear relationship with the subjective acceptance ratios, thus indicating the feasibility of using FQPath for slide-level quality control.

\section{Conclusion}\label{conclusion}
This paper presented an accurate and computationally efficient automated no-reference focus quality assessment (NR-FQA) technique that is currently needed for digital pathology workflows. It provides a practical solution for high-throughput scanning tasks that can be utilized in both image patch (aka block) and whole slide image (WSI) levels for focus quality evaluation of image scans. The method estimates the point spread function corresponding to the out-of-focus of the scanner imaging optics and synthesizes its inverse response as a sum of even-derivative filter bases. This kernel behaves similarly to the human visual system by modifying attenuated high-frequency image information. A set of digital pathology images are convolved with this synthesized filter to extract focus quality-related features and quantified these features to determine the quality score. The accuracy and computational efficiency of the proposed method are demonstrated in two fold: (a) the metric is evaluated at patch level and compared to ten other state-of-the-art NR-FQA metrics; and (b) the application of the proposed metric is demonstrated at WSI level to generate local focus quality maps (heatmaps), which are showed to be usable for automatically quantifying a slide's overall focus quality. 

The proposed NR-FQA metric retains both accuracy and speed that can favorably extended in different medical imaging applications for better engineering of QC control in helping clinicians/physicians to better serve the public health. Modalities include but not limited to brightfield microscopy, darkfield microscopy, fluorescence microscopy, confocal microscopy, etc. The metric could also be used to analyze the relevance of out-of-focus for developing automated diagnosis tools that have recently gained attention in computational pathology \cite{sari2018unsupervised, mercan2018multi, marsh2018deep}.

\section{Acknowledgment}
The authors would like to greatly thank Huron Digital Pathology, for providing valuable discussion and a database for developing the experiments conducted in this paper. The first, second, and third authors' research was partially supported by an NSERC Collaborative Research and Development Grant (contract CRDPJ-486583-15).

\ifCLASSOPTIONcaptionsoff
  \newpage
\fi

\bibliographystyle{IEEEbib}
\bibliography{refs}

\end{document}